\documentclass[12pt,showpacs]{revtex4-1}%

\usepackage[english]{babel}
\usepackage[T1]{fontenc}
\usepackage{bbm}
\usepackage{epsfig}
\usepackage{color}
\usepackage{amsmath}
\usepackage{amsfonts}
\usepackage{amssymb}
\usepackage{graphicx}%
\newcommand{\be}{\begin{equation}}
\newcommand{\ee}{\end{equation}}
\newcommand{\bea}{\begin{eqnarray}}
\newcommand{\eea}{\end{eqnarray}}

\renewcommand{\theequation}{\arabic{section}.\arabic{equation}}

\begin{document}

\title{Hairy black holes in the ghost-free bigravity theory
}

\author{Mikhail~S.~Volkov}

\affiliation{
Laboratoire de Math\'{e}matiques et Physique Th\'{e}orique CNRS-UMR 6083, 
Universit\'{e} de Tours, Parc de Grandmont, 37200 Tours, FRANCE}

\email{volkov@lmpt.univ-tours.fr}

\begin{abstract}
We study 
black holes in the recently proposed ghost-free theory 
with two gravitons, one of which is massive and another is massless. 
These black holes possess a regular event horizon 
which is common for both metrics and has the same 
values of the surface gravity and 
Hawking temperature with respect to 
each metric. The ratio of the event horizon radii measured by 
 the two metrics is a free parameter that labels 
the solutions. We present a numerical evidence for their existence 
and find that they   
comprise several classes.    
Black holes within each class 
approach the same AdS-type asymptotic  at infinity but differ 
from each other in the event horizon vicinity where the 
short-range massive modes reside. 
In addition, there are solutions showing  
a curvature singularity at a finite proper distance from the horizon.
For some special solutions the graviton mass may become effectively imaginary,
causing oscillations around the flat metric at infinity.  
The only asymptotically flat black hole we find -- the   
Schwarzschild solution obtained by identifying 
the two metrics --  seems to be exceptional, since 
 changing even slightly its horizon boundary conditions 
completely changes the asymptotic behavior at infinity. 
We also construct globally regular  solutions 
describing `lumps of pure gravity' which can be viewed as black hole remnants 
in the limit where the event horizon shrinks. 
Finally, adding a matter source we obtain globally regular and  
asymptotically flat solutions exhibiting the Vainstein mechanism
of recovery of General Relativity in a finite region.

\end{abstract}



\pacs{04.50.-h,04.50.Kd,04.70.Bw,04.25.dg}
\maketitle



\section{Introduction}
\setcounter{equation}{0}

The theoretic possibility to give a non-zero mass to gravitons \cite{Pauli} was considered  for 
a long time to have merely an academic interest. However, the 
recent observational evidence \cite{Reiss} suggests taking this idea more seriously,
because it could provide an explanation for the current acceleration
of our universe \cite{DK}. This has produced an increase of interest 
towards massive gravity theories (see \cite{Rubakov} for a recent review). 

Such theories are known to have serious theoretical 
difficulties -- the absence of a smooth 
massless limit \cite{vDVZ}, presence of the ghost state in the spectrum
\cite{BD}, and the very low ultraviolet cutoff \cite{AH}. 
However,
it seems that remedies may exist for some or perhaps for all of these
problems. For example, the van-Dam-Veltman-Zakharov (vDVZ) 
discontinuity in the massless 
limit \cite{vDVZ} seems to be curable by the Vainstein mechanism 
\cite{Vainstein}, as was recently
confirmed by the explicit calculation \cite{Babichev}. 
In addition,  it seems that the
presence of the Boulevard-Deser 
ghost and the absence of uniqueness may cure each other, 
since among many possible theories of massive gravity there could be one that
is ghost-free. 
 
A special massive gravity model  which is the only one
that is ghost-free in the decoupling limit was recently discovered
by de Rham, Gabadadze, and Tolley (RGT)
\cite{RGT}. 
Soon after this it was demonstrated that the model should in fact be  
completely free of the ghost \cite{HR},
which conclusion was independently confirmed by several other groups \cite{noghost}.
Therefore, although the opposite claim has been advocated as well \cite{ghost}, 
it seems that the RGT model of \cite{RGT}
is indeed completely free of non-physical ghost states.  
This property  manifests itself already at the level of the
equations of motion, 
some of which 
`miraculously' become algebraic, instead of being differential. 
The model has therefore received a lot of attention, in particular
its static solutions \cite{static}, \cite{statica} and time-dependent 
cosmological solutions \cite{cosm} have been studied. 

The RGT model is a theory of a massive graviton described by 
the dynamical metric $g_{\mu\nu}$ and 
the non-dynamical flat reference metric $f_{\mu\nu}$. 
There exist also {\it bigravity} theories 
suggested long ago by Isham, Salam and Strathdee \cite{ISS}, in which 
both  metrics are dynamical and 
describe two gravitons, one of which is massive and another is massless.
There can be many such theories, since the coupling between the two 
metrics can be arbitrary, with the only requirement that in the weak field limit 
it should assume the Pauli-Fierz form \cite{Pauli}. The latter insures that 
the number of propagating states in the linearized theory matches the number of  
polarizations of the two graviton. 
However, at the full non-linear level the generic bigravity develops  
additional non-physical propagating states due to the Boulevard-Deser ghost \cite{BD}.

To get rid of the latter, one can try to choose the interaction
between the two metrics in a special way, and the recipe for this was 
recently given in Ref.\cite{HR1}:
 to choose  the same interaction as in the RGT model (see Eq.\eqref{lagr} below). 
In other words, it was suggested to promote the RGT model to the bigravity by adding
the kinetic term for the second metric.  
The analysis of \cite{HR2} has then confirmed that the resulting theory is indeed
ghost-free, so that it provides an example of a bigravity without
unphysical states. 
This theory contains the RGT model as a special limit,
but it has richer dynamics, since it contains more degrees of freedom. 
For example, it reproduces General Relativity when the two
metrics are identified, whereas otherwise the massive 
degrees of freedom are present. In this respect the theory resembles 
the standard gauge field theories, where the massive states are generated  
by assuming a non-zero value for the Higgs field, 
but one can just as well switch  off the mass by setting the Higgs field to zero.    
Cosmological solutions in this theory  were studied in 
\cite{cosm1}, \cite{cosm1a} 
while some static solutions were obtained in \cite{static1}. 

In the present paper we carry out a detailed analysis of 
static, spherically symmetric solutions
in the ghost-free bigravity of \cite{HR1}, manly focusing on black holes. 
Our motivation is provided by the fact that all known black holes
in massive gravity/bigravity theories are essentially of the same special type, 
obtained within a special field ansatz. 
They were originally discovered 
in a particular bigravity model by Isham and Storey \cite{Isham}, but later similar 
solutions were found in other theories
\cite{static}, \cite{statica}.
For these solutions the metrics $g_{\mu\nu}$ and $f_{\mu\nu}$ are chosen to be static and 
not simultaneously diagonal, in which case the off-diagonal components of the Einstein 
equations impose a strong constraint which restricts  
the metric $g_{\mu\nu}$ to be Schwarzschild-(anti)de Sitter,
possibly with additional power-law corrections
 \cite{staticb}, \cite{staticc}. 
It is however unclear if other,
more general, asymptotically flat, say, solutions can exist.

The two metrics can also be chosen to be simultaneously diagonal, in which case
the off-diagonal constraint does not arise and the resulting field equations 
turn out to be rather complex. 
They admit asymptotically flat solutions when a regular matter source 
is included  \cite{Babichev}, but it is unclear if they possess 
black hole solutions as well. The recent results 
seem to exclude such a possibility at least in the  massive gravity models where
one of the metrics is flat. 
Specifically, it turns out that when one
numerically integrates the field equations starting from 
infinity towards the inner region, one does not find a regular event horizon
but a naked singularity instead 
\cite{gruzinov}. In addition, the analysis of \cite{DJ} shows
that if a regular horizon is present, then it should be common for both 
metrics, which is impossible if one of them is flat.

At the same time,  asymptotically flat black holes
certainly exist in the bigravity theory, where both metrics 
are dynamical.  These are the usual vacuum black holes obtained by 
setting $g_{\mu\nu}=f_{\mu\nu}$. 
It is then natural to try to find their generalizations 
to the case where $g_{\mu\nu}\neq f_{\mu\nu}$ and 
the massive degrees of freedom
are excited. Such solutions
could describe black holes
surrounded by a cloud of massive hair. 
We therefore study in what follows black holes 
in the ghost-free bigravity of \cite{HR1}, mainly focusing on
the case where both 
metrics are diagonal, the opposite  case being considered in the Appendix.

We assume a regular even horizon and starting from it 
integrate numerically the 
field equations 
towards infinity. 
We find  that the horizon is common for both metrics
and that the values of its surface gravity and 
Hawking temperature with respect to 
each metric are the same, which agrees with the conclusions of 
Refs.\cite{staticc}, \cite{DJ}. 
The ratio of the event horizon radii measured by 
 the two metrics is a free parameter that labels 
the solutions, so it is sufficient to vary only
this parameter to construct all possible black holes.  
We discover in this way that the generic solutions   
approach at infinity either the anti-de Sitter (AdS) metric,  or 
special metrics that we call $U,a$ backgrounds (see Eq.\eqref{Ua} below),
or they develop a curvature singularity at a finite proper distance 
from the horizon. There are many solutions which run to the same 
AdS or $U,a$ asymptotic  at infinity but differ 
from each other in the event horizon vicinity where 
a short massive `hair' resides.  On the other hand, the asymptotically flat 
solution of Schwarzschild 
seems to be exceptional, since 
 changing even slightly its horizon boundary conditions
immediately changes the asymptotic behavior at infinity.  
As a result, we do find black holes with massive hair,
but they are not asymptotically flat. 
It seems that asymptotic flatness
is incompatible with the presence of a black hole horizon when the 
massive degrees of freedom are excited.

For the sake of completeness, 
we also study solutions without a horizon 
which describe `lumps of pure gravity'. 
They have a regular center, but they are also not asymptotically flat 
and show in the far field the same behavior as the black holes. 
It seems they can be obtained from the latter in the limit 
where the horizon shrinks to zero.  
However, adding a matter source we  do find   
asymptotically flat solutions which exhibit the Vainstein
mechanism of 
recovery of General Relativity at finite distances. 
This suggests  that the mechanism  needs a matter source
and so does not work for pure vacuum systems like  black holes.

In the following three sections we describe the ghost-free bigravity of \cite{HR1},
its equations of motion, the reduction to the spherically symmetric sector, 
and simple exact solutions that we call `background black holes'
and `$U,a$ backgrounds'.  
Sec.V explains our procedure of numerical integration of the 
equations starting from the horizon. In Sec.VI
we study solutions which are parametrically close to the analytically known 
background black holes. The generic parameter values are considered in Sec.VII,
with some limiting cases 
analyzed in Sec.VIII. In Sec.IX we present the globally regular 
solutions, while Sec.X. contains concluding remarks. 
Finally, black holes with   
non-simultaneously diagonal metrics are described in the Appendix. 

Our conventions and notation follow those of Ref.\cite{cosm1}, 
up to replacing $\eta\to\tan^2\eta$ and $m\to m\cos\eta$.

\section{The ghost-free bigravity  \label{RGT}}

\setcounter{equation}{0}

The generic bigravity theory 
is defined on  a four-dimensional spacetime manifold spanned by 
coordinates $x^\mu$ and equipped with two metrics $g_{\mu\nu}(x)$ and  $f_{\mu\nu}(x)$
whose kinetic terms are chosen to be of the 
standard Einstein-Hilbert form.
The action is \cite{ISS}
\be
S=-\frac{1}{16\pi G}\,\int R\sqrt{-g}\,d^4x-\frac{1}{16\pi {\cal G}}\,\int {\cal R}\,\sqrt{-f}d^4x
+S_{\rm int }[g_{\mu\nu},f_{\mu\nu}]
+S_{\rm m}[g_{\mu\nu},{\rm matter}]\,,
\ee
where $R$ and ${\cal R}$ are the Ricci scalars for $g_{\mu\nu}$ and $f_{\mu\nu}$, 
respectively, $G$ and ${\cal G}$ are the corresponding gravitational couplings,
while $S_{\rm m}$ describes ordinary matter
(for example perfect fluid) which is supposed  to directly interact only with 
$g_{\mu\nu}$. The interaction between the two metrics,
$S_{\rm int}$, 
remains largely arbitrary in the generic case. The only condition is that in the weak field limit
it should reduce to the Pauli-Fierz expression \cite{Pauli}, to insure that the number 
of propagating degrees of freedom in the linearized theory matches the number of the graviton polarizations.
This condition specifies only the structure of the quadratic part of the interaction term. 

When the fields are not
weak, the generic bigravity  develops, apart from the graviton degrees of freedom,  
an additional propagating mode in the spectrum -- the Boulevard-Deser ghost \cite{BD}. 
This mode is unphysical and should be excluded. According to 
Refs.\cite{RGT}, \cite{HR1}, it will be absent if  
the interaction 
is chosen as 
\be                                \label{int}
S_{\rm int}=\frac{\sigma}{8\pi G}\int {\cal L}_{\rm int}\sqrt{-g}\, d^4x 
\ee
where $\sigma$ is a parameter and 
\bea                                           \label{lagr}
\mathcal{L}_{\rm int}=\frac{1}{2}((K^\mu_\mu)^2-K_{\mu}^{\nu}K_{\nu}^{\mu})
+\frac{c_{3}}{3!}%
\,\epsilon_{\mu\nu\rho\sigma}
\epsilon^{\alpha\beta\gamma\sigma}K_{\alpha}^{\mu}
K_{\beta}^{\nu}K_{\gamma}^{\rho}+\frac{c_{4}}%
{4!}\,
\epsilon_{\mu\nu\rho\sigma}
\epsilon^{\alpha\beta\gamma\delta}K_{\alpha}^{\mu}
K_{\beta}^{\nu}K_{\gamma}^{\rho}K_{\delta}^{\sigma}\,.
\eea
Here 
$K^\mu_\nu=\delta^\mu_\nu-\gamma^\mu_{~\nu}$
where $\gamma^\nu_{~\nu}$ is defined 
by the relation 
\be                                             \label{gam0}
\gamma^\mu_{~\sigma}\gamma^\sigma_{~\nu}=g^{\mu\sigma}f_{\sigma\nu}
\ee
with $g^{\mu\nu}$ being the inverse of $g_{\mu\nu}$, while  $c_3,c_4$ are free parameters.  
These expressions define the ghost-free theory with two gravitons, one of which is massless and
another one is massive. 
The mass of the latter is given by
\be                             \label{m}
m^2=\sigma\left(1+\frac{\cal G}{G}\right).
\ee
In the limit where 
${\cal G}\to 0$ and $f_{\mu\nu}$ is flat, the theory reduces to 
the RGT model of \cite{RGT}. An equivalent parameterization 
of the theory in terms of eigenvalues of $\gamma^\sigma_{~\nu}$ is often
considered in the literature \cite{HR1}, while our parameterization \eqref{lagr}
agrees with that of Ref.\cite{statica}.  

One can introduce an angle $\eta$ such that the parameters $\sigma,{\cal G}$
subject to the condition \eqref{m} are expressed according to one of the following 
three possibilities,   
\begin{subequations}                
\begin{align}
\sigma&=m^2\cos^2\eta,~~~~~~~~~~~{\cal G}=G\tan^2\eta ; \label{al1} \\
\sigma&=m^2\cosh^2\eta,~~~~~~~~~{\cal G}=-G\tanh^2\eta; \label{al2}\\
\sigma&=-m^2\sinh^2\eta,~~~~~~~{\cal G}=-G\coth^2\eta.  \label{al3}
\end{align}
\end{subequations}
For most of our discussion below we shall assume the first
of these options, in which case both $\sigma$ and ${\cal G}$ are positive. 
The other two options will be briefly discussed in Section VII. 

The tensor $\gamma^\mu_{~\sigma}$ defined by \eqref{gam0}
is hard to explicitly express in terms of 
$g_{\mu\sigma}$ and $f_{\sigma\nu}$,  
which creates difficulties when varying the action \cite{cosm1}. 
To handle the problem, it is very convenient to 
introduce two  tetrads
$e_A^\mu$ and $\omega^A_\mu$ defined by the conditions 
$g^{\mu\nu}=\eta^{AB}e_A^\mu e_B^\nu$ 
and
$f_{\mu\nu}=\eta_{AB}\omega^A_\mu \omega^B_\nu$
with $\eta_{AB}={\rm diag}(1,-1,-1,-1)$. The tetrads 
are defined up to the local $SL(1,3)\times SL(1,3)$ rotations,
which freedom can be used to impose the conditions 
\be                       \label{const}
e_A^\mu\omega_{B\mu}=e_B^\mu\omega_{A\mu}\,
\ee
with $\omega_{A\mu}=\eta_{AB}\omega^B_\mu$
which insures that 
\be                      \label{gam}
\gamma^\mu_{~\nu}=e^\mu_A\omega^A_\nu\,.
\ee
Using this explicit representation, 
the action can be varied with respect to $e^\mu_A$ and $\omega^A_\nu$
(see \cite{cosm1} for details), after which the resulting field equations 
assume the form
\begin{align}
G^\rho_\lambda&=m^2\cos^2\eta\, T^\rho_\lambda
+8\pi G T^{{\rm (m)}\,\rho}_{~~~~\lambda} \,, \label{e1} \\
{\cal G}^\rho_\lambda&=
m^2\sin^2\eta\, {\cal T}^\rho_\lambda    \,.        \label{e2}
\end{align}
Here $G^\rho_\lambda$ and ${\cal G}^\rho_\lambda$ are the Einstein tensors 
for $g_{\mu\nu}$ and $f_{\mu\nu}$, 
respectively, while 
\be                                   \label{TTT0}
T^\rho_\lambda=\tau^\rho_\lambda
-\delta^\rho_\lambda\,{\cal L}_{\rm int}\,,~~~~~~~
{\cal T}^\rho_\lambda=-\frac{\sqrt{-g}}{\sqrt{-f}}\, 
\tau^\rho_\lambda\,,
\ee
with 
\be					\label{tau}
\tau^\rho_\lambda=(\gamma^\sigma_\sigma-3)\gamma^\rho_\lambda
-
\gamma^\rho_\sigma\gamma^\sigma_\lambda 
-\frac{c_{3}}{2}%
\,\epsilon_{\lambda\mu\nu\sigma}
\epsilon^{\alpha\beta\gamma\sigma}\gamma_{\alpha}^{\rho}
K_{\beta}^{\mu}K_{\gamma}^{\nu}
-\frac{c_{4}}{6}\,
\epsilon_{\lambda\mu\nu\sigma}
\epsilon^{\alpha\beta\gamma\delta}
\gamma_{\alpha}^{\rho}
K_{\beta}^{\mu}K_{\gamma}^{\nu}K_{\delta}^{\sigma}\,.  
\ee
The Bianchi identities for the left-hand side of Eq.\eqref{e1} imply
the conservation condition 
\be                                   \label{T1} 
\stackrel{(g)}{\nabla}_\rho T^\rho_\lambda=0\,,
\ee
where $\stackrel{(g)}{\nabla}_\rho$ is the covariant derivative with respect to  
$g_{\mu\nu}$. The conditions $\stackrel{(f)}{\nabla}_\rho {\cal T}^\rho_\lambda=0$
follows from \eqref{T1} in view 
of the diffeomorphism-invariance of the interaction term 
$S_{\rm int}$
 (see \cite{cosm1} for details). 
The matter energy-momentum tensor is conserved independently, in view 
of the diffeomorphism-invariance of the matter action $S_{\rm m}$,
\be                          \label{T2}
\stackrel{(g)}{\nabla}_\rho T^{{\rm (m)}\rho}_{~~~~\lambda}=0\,.
\ee
If the matter source vanishes, then 
choosing the two metrics to be the same reduces the field equations 
to those of vacuum General Relativity,
\be
g_{\mu\nu}=f_{\mu\nu},~~~~
T^{{\rm (m)}\rho}_{~~~~\lambda}=0
~~~~~~~\Rightarrow~~~~~~ 
G^\rho_\lambda={\cal G}^\rho_\lambda=0,
\ee
since one has in this case $T^\mu_\nu={\cal T}^\mu_\nu=0$. 

In what follows we shall be considering solutions of 
equations \eqref{e1}, \eqref{e2} within the spherically symmetric sector.  
Introducing the spherical coordinates 
$x^\mu=(t,r,\vartheta,\varphi)$, the 
most general expressions for the two tetrads are \cite{cosm1}
\bea               
e_0&=&\frac{1}{Q}\,\frac{\partial}{\partial t },~~~~~~~~~
e_1={N}\,\frac{\partial}{\partial r },~~~~~~~~~
e_2=\frac{1}{R}\,\frac{\partial}{\partial \theta },~~~~~~~~~
e_3=\frac{1}{R\sin\vartheta}\,\frac{\partial}{\partial \varphi },~~~~\notag \\
\omega^0&=&a\,dt+c\,dr,~~~~\omega^1=-{c\,QN}\,dt+b\,dr,
~~~~\omega^2=Ud\vartheta,
~~~~\omega^3=U\sin\vartheta d\varphi\,,                    \label{tetrad}
\eea 
where $Q,N,R,a,b,c,U$ are functions of $r$. 
The corresponding metrics read 
\be
g_{\mu\nu}dx^\mu dx^\nu=Q^2dt^2-\frac{dr^2}{N^2}-R^2(d\vartheta^2+\sin^2\vartheta d\varphi^2)
\ee
and
\be                             \label{fff}
f_{\mu\nu}dx^\mu dx^\nu=(a^2-c^2Q^2N^2)\,dt^2+2{c(a+bQN)}\,dtdr
-(b^2-c^2)\, dr^2-U^2(d\vartheta^2+\sin^2\vartheta d\varphi^2),
\ee
while 
\be                                  \label{gamma}
\gamma^\mu_{~\nu}=e_A^\mu \omega^A_\nu=\left(
\begin{array}{cccc}
{a}/{Q} & {c}/{Q} & 0 & 0 \\
-{c\,QN^2} & {bN} & 0 & 0 \\
0 & 0 & {U}/{R} & 0 \\
0 & 0 & 0 & {U}/{R}
\end{array}
\right)\,.
\ee
It is now straightforward to compute 
${\cal L}_{\rm int}$ and the tensor $\tau^\mu_\nu$ defined by \eqref{tau}
(the explicit expressions are given in the Appendix in Ref.\cite{cosm1}).

 It is still possible to reparameterize the radial 
coordinate to impose the 
gauge condition $R(r)=r$ so that 
 the metric $g_{\mu\nu}$ becomes 
\be                                \label{ggg}
g_{\mu\nu}dx^\mu dx^\nu=Q^2 dt^2-\frac{dr^2}{N^2}
-r^2(d\vartheta^2+\sin^2\vartheta d\varphi^2).
\ee
Since its Einstein tensor is diagonal, so should be the
energy-momentum tensor. Therefore, 
one has to have $T^0_r= 0$,
which requires that $\tau^0_r= 0$ (this is the `off-diagonal constraint'
mentioned in the Introduction). 
Using \eqref{tau} one finds  
\bea                                 \label{T0r}
\tau^0_r=\frac{c}{Qr^2}
\left( r\,(2U-3r)+ c_3\,(3r-U)(r-U)
+c_4\,(r-U)^2 \right),
\eea
and for this to vanish, one should either have $c=0$, or 
choose $c\neq 0$ but set to zero the 
expression between the parenthesis. 
The latter case is studied 
in the Appendix, where it is shown that the metric $g_{\mu\nu}$ 
will be  described
in this situation  by the  
Schwarzschild-(anti)de Sitter solution. More general metrics are obtained 
by choosing $c=0$.

\section{Field equations}
\setcounter{equation}{0}

If $c=0$ then the metric $f_{\mu\nu}$ in \eqref{fff} becomes diagonal. 
Assuming that all its coefficient depend only on $r$ and 
introducing a new function $Y$ via
$b=U^\prime/Y$ with $^\prime\equiv d/dr$ the metric reads
\be                                \label{ggg1}
f_{\mu\nu}dx^\mu dx^\nu=a^2 dt^2-\frac{U^{\prime 2}}{Y^2}\,dr^2
-U^2(d\vartheta^2+\sin^2\vartheta d\varphi^2). 
\ee
We thus have two static and spherically symmetric metrics 
\eqref{ggg} and \eqref{ggg1}
which contain 5 functions of $r$: $Q$, $N$, $a$, $Y$, $U$. 
The Einstein equations \eqref{e1},\eqref{e2} reduce to
\bea                              \notag
G^0_0&=&{m^2\cos^2\eta\,} T^0_0+\rho, \notag \\
G^r_r&=&{m^2\cos^2\eta\,} T^r_r-P,~~~\notag \\
{\cal G}^0_0&=&{m^2\sin^2\eta\,} {\cal T}^0_0,\notag \\
{\cal G}^r_r&=&{m^2\sin^2\eta\,} {\cal T}^r_r,\label{GGG}
\eea
which should be supplemented by the 
conservation condition 
$\stackrel{(g)}{\nabla}_\rho T^\rho_\lambda=0$, 
\bea 
(T^r_r)^\prime &+&\frac{Q^\prime}{Q}\,(T^r_r-T^0_0)
+\frac{2}{r}(T^\vartheta_\vartheta-T^r_r)=0.   \label{TTT}
\eea
As was said above, it is not necessary to require in addition that 
$\stackrel{(f)}{\nabla}_\rho {\cal T}^\rho_\lambda=0$, since this condition 
follows from \eqref{GGG},\eqref{TTT}. It is assumed in the above 
equations that the matter source is of the perfect fluid type, 
\be
8\pi G T^{{\rm (m)}\,\rho}_{~~~~\lambda}={\rm diag}[\rho(r),-P(r),-P(r),-P(r)]
\ee
whose conservation requires that 
\be                       \label{pressure} 
P^\prime=-\frac{Q^\prime}{Q}\,(\rho+P).
\ee
The 5 equations \eqref{GGG},\eqref{TTT} explicitly read
\begin{subequations}                
\begin{align}                    
\frac{ 2NN^\prime}{r}+\frac{N^2-1}{r^2}+m^2\cos^2\eta
\left(\alpha_1\,\frac{N}{Y}\,U^\prime
+\alpha_2\right)+\rho&=0,   \label{eq1} \\
\frac{ 2N^2Q^\prime}{Qr}+\frac{N^2-1}{r^2}+m^2\cos^2\eta
\left(\alpha_1\,\frac{a}{Q}+\alpha_2
\right)-P&=0,   \label{eq2} \\
\{Y^2-1+m^2\sin^2\eta\,\alpha_3\}NU^\prime+2UNYY^\prime
+m^2\sin^2\eta\,Y\alpha_4&=0, \label{eq3}     \\
\{a(Y^2-1)+m^2\sin^2\eta\,\alpha_5\}U^\prime+2UY^2a^\prime&=0,  \label{eq4}  \\
\alpha_6U^\prime +\alpha_7a^\prime&=0,  \label{eq5}
\end{align}
\end{subequations}
where the following abbreviations have been introduced,
\bea
\alpha_1&=&
3-3 c_3-c_4+\frac{2(c_4+2c_3-1)U}{r}
-\frac{(c_4+c_3)U^2}{r^2},~~~~~\notag \\
\alpha_2&=&4c_3+c_4-6+\frac{2(3-c_4-3c_3)U}{r}
+\frac{(c_4+2c_3-1)U^2}{r^2}, \notag \\
\alpha_3&=&c_4U^2-2(c_3+c_4)rU+(c_4+2c_3-1)r^2,\notag \\
\alpha_4&=&(3-c_4-3c_3)r^2-(c_4+c_3)U^2+(4c_3+2c_4-2)rU ,\notag \\
\alpha_5&=&[(a-Q)c_4-Qc_3]U^2
+[2(2Q-a)c_3+(Q-a)c_4-Q]rU ,\notag \\
&+&[(2a-3Q)c_3+(a-Q)c_4+3Q-a]r^2 , \notag  \\
\alpha_6&=&Q^\prime N[(3c_3+c_4-3)r^2
+(2(1-c_4-2c_3))Ur
+(c_4+c_3)U^2] ,  \notag \\
&+&2Q(Y-N)[(3-c_4-3c_3)r
+(c_4+2c_3-1)U] , \notag \\
&+&2a(N-Y)[
(1-c_4-2c_3)r
+(c_4+c_3)U],\notag \\
\alpha_7&=&
Y[(3-c_4-3c_3)r^2+2(c_4+2c3-1)Ur-(c_4+c_3)U^2].  
\eea
Eliminating the term $Q^\prime$ in $\alpha_6$ 
with the use of   
\eqref{eq2}, the coefficients $\alpha_1,\ldots,\alpha_7$  
depend only on $Q$, $N$, $a$, $Y$, $U$ but
not on their derivatives. 

The equations admit the scale symmetry that maps solutions
to solutions, 
\bea                                   \label{scale}
N(r)&\to& N(\lambda r),~~~~Y(r)\to Y(\lambda r),~~~~
U(r)\to \frac{1}{\lambda}\,U(\lambda r),~~~~
Q(r)\to Q(\lambda r),\notag \\
a(r)&\to& a(\lambda r),~~~~ 
m\to\frac{m}{\lambda},~~~~\eta\to\eta,~~~~c_3\to c_3,~~~~c_4\to c_4. 
\eea
In the following few sections, until Section IX, we shall set $\rho=P=0$.

\section{Simplest solutions}
\setcounter{equation}{0}

Some exact solutions of equations \eqref{eq1}--\eqref{eq5}
can be obtained.

\subsection{Background black holes}

Let us choose the two metrics to be conformally related,  
\be
f_{\mu\nu}=C^2g_{\mu\nu},
\ee
with constant $C$. Equations \eqref{eq1}-\eqref{eq5}
will be fulfilled if $C$ satisfies the algebraic equation 
\be                                \label{b0}
(C-1)P(C)=0,
\ee
where (with $\xi=\tan^2\eta$)
\be                                   \label{b3}
P(C)=(c_3+c_4)C^2+(3-5c_3+(\xi-2)c_4)C+(4-3\xi)c_3+(1-2\xi)c_4-6
+\frac{\xi(3c_3+c_4-3)}{C}
\ee
while $g_{\mu\nu}$ is Schwarzschild-(anti)de Sitter,  so that
\be                                 \label{b1}
Y^2=N^2=1-\frac{2M}{r}-\frac{\Lambda(C)}{3}\,r^2,~~~~
U=Cr,~~~~~a=CQ,~~~Q=qN,
\ee                                      
where $q$ is an arbitrary constant related to the time scaling symmetry and
\be                                         \label{b00}
\Lambda(C)=m^2\cos^2\eta(1-C)\{(c_3+c_4)C^2+(3-5c_3-2c_4)C+4c_3+c_4-6 \}.
\ee
One root of Eq.\eqref{b0} is $C=1$, 
in which case $\Lambda=0$ and we obtain the Schwarzschild solution.
Depending on values of the parameters $c_3$, $c_4$, $\eta$, equation \eqref{b0}
can have up to three more real roots, for which $\Lambda(C)$ 
does not generically vanish.  
For example, for
$\eta=1$, $c_3=0.1$, $c_4=0.3$ Eq.\eqref{b0} has 
altogether four 
real roots, $C=\{C_k\}$, 
\bea                           \label{roots}
\{C_1,C_2,C_3,C_4\}&=&\{1;\,-0.6458\,;2.6333\,;-8.5566\}, \notag \\
\frac{\Lambda(C_k)}{m^2}&=&\{0;\,-3.0559;\,-1.1812;\,+21.5625\}. 
\eea
This gives the Schwarzschild (S) solution for $C=C_1$, 
Schwarzschild-anti-de Sitter (SAdS) solutions for $C=C_2,C_3$, and 
the  Schwarzschild-de Sitter (SdS) solution for $C=C_4$. We shall call these 
solutions background black holes, since below they will be considered as 
reference backgrounds -- the starting point for studying more general solutions.  

It is worth noting that these 
 solutions are not the same as those described in 
the Appendix, where $g_{\mu\nu}$ is also 
SdS or SAdS but $g_{\mu\nu}$ and $f_{\mu\nu}$ are not proportional.


\subsection{$U,a$ backgrounds}

Another class of solutions can be obtained by 
setting $U,a$ to constant values:
\bea                                        \label{Ua}
N^2&=&1+m^2\cos^2\eta \left((1-2c_3-c_4)U^2-\frac{2M}{r}+(3c_3+c_4-3)Ur
+(2-\frac{4}{3}\,c_3-\frac{1}{3}\,c_4)r^2\right),  \notag \\
{\frac{Q}{N}}&=&a\frac{m^2\cos^2\eta}{2}\int_{r_1}^r \frac{dr}{x{N}^{3}}\,
{\cal F},~~~~~~~~~~~~
Y=\frac{m^2\sin^2\eta}{2U}\int^r_{r_2} \frac{dr}{{N}}\,
{\cal F},
\eea
where
${\cal F}=(c_4-3+3c_3)x^2+2(1-2c_3-c_4)Ux+(c_3+c_4)U^2$ 
and $M$, $r_1$, $r_2$ are integration constants. 
For $r\to\infty$ one has $Q^2\sim N^2\sim Y\sim r^2$ 
and the metric $g_{\mu\nu}$ approaches in the leading order 
the (anti)de Sitter metric, but
the subleading terms are different. 
The metric $f_{\mu\nu}$ is actually degenerate,
since $f_{rr}=U^{\prime 2}/Y^2=0$. However, such 
solutions will describe below the asymptotic behavior of other, 
more general solutions for which 
$U,a$ become constant only for $r\to\infty$
so that $f_{rr}$ vanishes
only asymptotically. The proper distance up to infinity 
$\int^\infty_r (U^\prime/Y) dr$ and the proper volume of the 3-space are then 
finite, so that with respect to $f_{\mu\nu}$ the spacetime is spontaneously 
compactified.

\section{Boundary conditions at the horizon}
\setcounter{equation}{0}
Since we could not find other solutions of equations \eqref{eq1}--\eqref{eq5}
in a closed analytic form, we wish to integrate these equations numerically. 
As a first step, one should resolve the equations with respect 
to the derivatives, and our procedure is as follows. 
First of all, 
taking the ratio of Eqs.\eqref{eq4} and \eqref{eq5} yields  the algebraic relation
\be
\frac{a(Y^2-1)+m^2\sin^2\eta\,\alpha_5}{\alpha_6}=\frac{2UY^2}{\alpha_7},  \label{eee}  
\ee
which can be expressed in the form 
\be                      \label{aaa}
\frac{a}{Q}={\cal F}(r,N,Y,U,m,\eta,c_3,c_4). 
\ee
Using this, Eqs.\eqref{eq1}, \eqref{eq2}, \eqref{eq3} and
\eqref{eq5} reduce to 
\begin{subequations}                
\begin{align}   
N^\prime&={\cal F}_1\,U^\prime+{\cal F}_2\,,\label{aaa1} \\
Y^\prime&={\cal F}_3\,U^\prime+{\cal F}_4\,,\label{aaa2} \\
Q^\prime&={\cal F}_5\,Q\,,\label{aaa3} \\
a^\prime&={\cal F}_6\, QU^\prime\,,\label{aaa4} 
\end{align}  
\end{subequations}                 
where ${\cal F}_k={\cal F}_k(r,N,Y,U,m,\eta,c_3,c_4)$. Eqs.\eqref{aaa},
\eqref{aaa3},\eqref{aaa4} together imply that
\be
{\cal F}^\prime={\cal F}_6\,U^\prime-{\cal F}{\cal F}_5\,,
\ee
and explicitly calculating the derivative on the left gives 
\be
\partial_r{\cal F}+\partial_N{\cal F}({\cal F}_1\,U^\prime+{\cal F}_2 )
+\partial_Y{\cal F}({\cal F}_3\,U^\prime+{\cal F}_4 )
+\partial_U{\cal F}U^\prime={\cal F}_6\,U^\prime-{\cal F}{\cal F}_5,
\ee
which can be resolved with respect to $U^\prime$,
\be
U^\prime=\frac{\partial_r{\cal F}+\partial_N{\cal F}{\cal F}_2 
+\partial_Y{\cal F}{\cal F}_4 +{\cal F}{\cal F}_5  }
{{\cal F}_6-{\cal F}_1\partial_N{\cal F}
-{\cal F}_3\partial_Y{\cal F}
-\partial_U{\cal F}
}\equiv {\cal D}U. 
\ee
Injecting this into \eqref{aaa1} and \eqref{aaa2} finally
yields a system of 3 coupled equations, 
\bea                            \label{eeqs}
N^\prime&=&{\cal D}N(r,N,Y,U,m,\eta,c_3,c_4),\notag \\
Y^\prime&=&{\cal D}Y(r,N,Y,U,m,\eta,c_3,c_4),\notag \\
U^\prime&=&{\cal D}U(r,N,Y,U,m,\eta,c_3,c_4), 
\eea
where the explicit expressions of the functions on the right 
are somewhat lengthy, so that we do not 
write them down explicitly. When a solution of these equations is found, then $Q$
is obtained by integrating equation \eqref{aaa3},  
and finally $a$ is obtained from \eqref{aaa}. 

We wish to study black hole solutions of Eqs.\eqref{eeqs} and so  assume that 
there is an event horizon at $r=r_h$ where $Q(r_h)=N(r_h)=0$. 
For the horizon to be non-singular and non-degenerate, $Q^2$ 
and $N^2$ should both have
simple zeros at  $r=r_h$. 
Next, the inspection  of Eqs.\eqref{eeqs}
shows that if $N^2$ has a simple zero at $r=r_h$  
then $Y^2$ should have a simple zero at this point too. 
At the same time, $U$ can assume 
at the horizon any finite value.  
We therefore assume 
the power-series expansions near $r=r_h$,
\be                      \label{local}
N^2=\sum_{n\geq 1}a_n(r-r_h)^n,~~~~~
Y^2=\sum_{n\geq 1}b_n(r-r_h)^n,~~~~~
U=ur_h+\sum_{n\geq 1}c_n(r-r_h)^n.
\ee
Injecting this to \eqref{eeqs}, 
all coefficients $a_n$, $b_n$, $c_n$ can be expressed in terms of $u,a_1$
where $u$ is arbitrary 
while  $a_1$ satisfies 
${\cal A}\,a_{1}^2+{\cal B}\,a_1+{\cal C}=0$, where ${\cal A}$, 
${\cal B}$, ${\cal C}$ are functions of $u,r_h,m,\eta,c_1,c_2$. 
There are two solutions for $a_1$,
\be                                     \label{a1}
a_{1}=a_{1}^{\pm}(u)=\frac{1}{2{\cal A}}\,(-{\cal B}\pm
\sqrt{{\cal B}^2-4{\cal A}{\cal C}  }), 
\ee
which gives rise to two different 
local solutions \eqref{local} with different values
of $(N^2)^\prime$, $(Y^2)^\prime$, $U^\prime$ at $r=r_h$.
Inserting these solutions into
Eqs.\eqref{aaa},\eqref{aaa3} yields 
\be                      \label{local1}
Q^2=q^2\{r-r_h+\sum_{n\geq 2}c_n(r-r_h)^n\},~~~~~
a^2=q^2\sum_{n\geq 1}d_n(r-r_h)^n,~~~~~
\ee
where $c_n$ and $d_n$ are expressed in terms of $a_1$ and $u$, while 
$q$ is an integration constant that reflects the possibility
to rescale the time coordinate for both metric simultaneously. 

At this point it is interesting to compare our results with the 
geometric analysis of bimetric theories of
Ref.\cite{DJ}. Let us consider $\xi=\partial/\partial t$, 
the timelike Killing vector
for the both metrics. 
Its norms calculated with respect to each 
metric, $\langle\xi,\xi\rangle_g=Q^2$ and $\langle\xi,\xi\rangle_f=a^2$,
both vanish at $r=r_h$ so that the horizon is the 
Killing horizon for each of the two metrics at the same time. 
The surface gravity for each metric is the 
horizon value of 
\bea
\kappa^2_{g}&=&-\frac12\,g^{\mu\alpha}g_{\nu\beta}
\stackrel{(g)}{\nabla}_\mu\xi^\nu\,  
\stackrel{(g)}{\nabla}_\alpha\xi^\beta=\lim_{r\to r_h}Q^2N^{\prime 2}
=\frac{1}{4}\, q^2a_1    \,,               \notag \\
\kappa^2_{f}&=&-\frac12\,f^{\mu\alpha}f_{\nu\beta}
\stackrel{(f)}{\nabla}_\mu\xi^\nu
\stackrel{(f)}{\nabla}_\alpha\xi^\beta=\lim_{r\to r_h}a^2
\left(\frac{Y}{U^\prime}\right)^{\prime 2}
=\frac{1}{4}\, q^2\,\frac{d_1 b_1}{(c_1)^2}\,.
\eea
The Hawking temperature for each metric 
is
obtained in the standard way by passing to the imaginary time 
and requiring the absence of conical singularity. This gives 
\be
T_g=\frac{\kappa_g}{2\pi}\,,~~~~~~T_f=\frac{\kappa_f}{2\pi}.
\ee 
Now, using the explicit expressions
for the coefficients $a_1$, $b_1$, $c_1$, $d_1$ (we do not 
write them down in view of their complexity) 
the ratio of the two surface gravities evaluates to one, 
\be
\frac{\kappa_g^2}{\kappa_f^2}=\frac{a_1(c_1)^2}{d_1 b_1}=1,
\ee 
for any $u$ and for both signs in \eqref{a1}. Therefore, the 
surface gravities and the Hawking temperatures are the same. 
All this agrees with the conclusions of Ref.\cite{DJ} (see also \cite{staticc})
that the two Killing horizons and their surface gravities should 
coincide.

Returning to our analysis, 
the black holes solutions are obtained by 
numerically extending the local solutions \eqref{local}, \eqref{local1} 
towards large $r$. 
It follows that the black holes
are determined by the value of  $u=U(r_h)/r_h$ --
the 
ratio of the event horizon
radius measured by $f_{\mu\nu}$ to that measured 
by $g_{\mu\nu}$ -- as well as by the choice of sign 
in \eqref{a1}. Depending on the latter, we shall say that the 
solutions belong either to the upper or to the 
lower branch.  
Under the scale transformations \eqref{scale} the ratio
$u=U(r_h)/r_h$ stays invariant while $r_h\to r_h/\lambda$, 
and this can be used to set 
$r_h=1$, which condition
will be assumed from now on.

Therefore, to explore the structure 
of the solution space, one has to integrate 
the equations starting from the horizon for various values 
of $u$, separately for each branch.  
The integration constant $q$ in \eqref{local1} is not 
an essential parameter and can be fixed afterwards, when the 
global solutions are already known, for example by requiring 
that $Q^2/N^2\to 1$ as $r\to\infty$.
Let us choose some values of the theory parameters, 
for example the same as in Eq.\eqref{roots}. 
There is nothing special about these values, since varying them changes 
the solutions smoothly, without changing  their qualitative structure. 
We then use the numerical routines of \cite{NR} to integrate 
the equations  starting from $r=r_h=1$ towards large 
$r$.

As a starting point, for $u=C$ with $C=\{C_k\}$ given by Eq.\eqref{roots}
the numerical solutions should reproduce the background black holes,
for which $U^\prime(r)=C$.  
And indeed, for the four values of $C$ in Eq.\eqref{roots}
we find numerical solutions with 
$U^\prime(r)=C$ and $N,Q,Y$ coinciding 
 with those in   Eq.\eqref{b1}. 
The solutions for $C=\{C_1,C_2\}$ belong to the upper branch,
while those for  $C=\{C_3,C_4\}$ belong to the lower branch.

\section{Hairy black holes} 
\setcounter{equation}{0}

As a next step, we choose $u=C+\delta u$ where $\delta u$ is small. 
It is then natural to expect the solution to be a slightly 
deformed 
background black hole, the deformations being due to the 
`massive hair' present for generic values of $u$.  
In what follows we analyze these expectations 
and find that 
the deformations are indeed small,
but in general only within a finite neighborhood of the event horizon 
and not necessarily for all values of $r$.  

\subsection{Deformations of the asymptotically flat black hole.}

The only asymptotically flat solution among the background black holes 
is the Schwarzschild solution \eqref{b1} with $C=C_1=1$. 
We consider
its deformed version for $u=1+\delta u$. It turns out that 
choosing $\delta u$ to be negative does not give anything, since 
 the local solutions \eqref{local} become 
then complex-valued. 
However, for $\delta u$ positive and small enough, for example 
$\delta u=10^{-2}$, the deformed solution exists  
and stays very close to the Schwarzschild solution in a large vicinity 
of the horizon, for $r<r_{\rm max}(u)$. 
However, for larger values of $r$ it 
completely changes its structure, since
 the $Q$, $N$, $Y$ amplitudes then grow rapidly,
while $a$, $U$ approach finite asymptotic values (see Fig.1),
so that the whole configuration approaches one of the 
$U,a$ backgrounds \eqref{Ua}. 

 \begin{figure}[th]
\hbox to \linewidth{ \hss
	
	\resizebox{8.5cm}{5.2cm}{\includegraphics{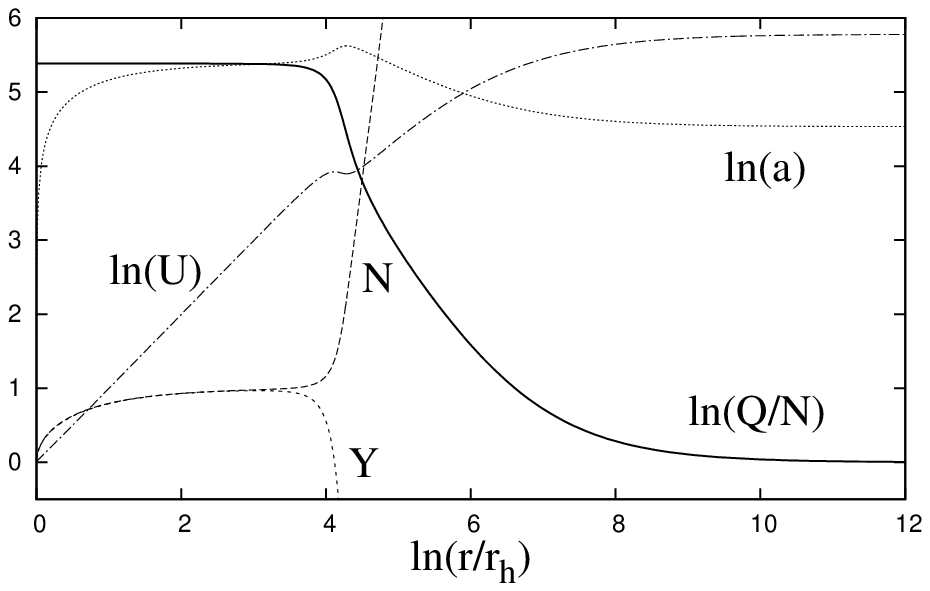}}
\hspace{1mm}
	\resizebox{8.5cm}{5.2cm}{\includegraphics{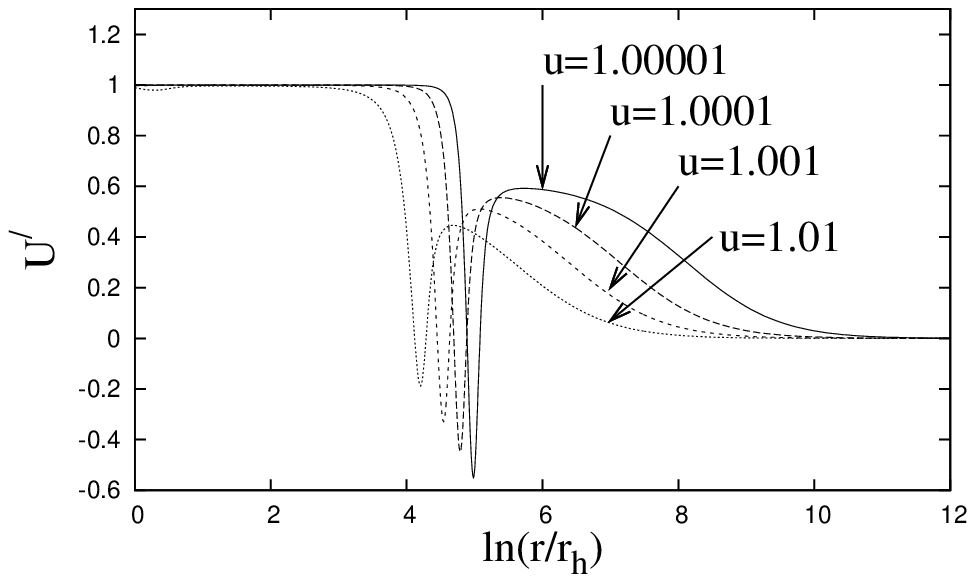}}
	
\hspace{1mm}
\hss}
\label{Fig1}
\caption{{\protect\small 
Left: solution profiles for $u=1.01$. Right: $U^\prime(r)$
for several values of $u$. Here 
and in Figs.2,3,4 below one has $m=0.1$, $\eta=1$, $c_3=0.1$, $c_4=0.3$.  
 }}%
\end{figure}

It is instructive to consider the function $U^\prime(r)$, which is equal 
to one everywhere for $u=1$. For $u>1$ it
stays very close to one for $r<r_{\rm max}(u)$ but for
$r\sim r_{\rm max}(u)$ it suddenly drops down and after a couple of 
oscillations tends to zero at infinity. The value $r_{\rm max}(u)$
increases when $u$ decreases, and in the limit 
$u\to 1$ one has $r_{\rm max}(u)\to\infty$ so that the solution 
approaches the Schwarzschild metric for any finite $r$. 
However, since  the boundary conditions at infinity for $u>1$ 
are not the same as for $u=1$,
the convergence in the limit $u\to 1$ is 
only pointwise and not uniform.

The conclusion is that exciting the massive 
degrees of freedom around the 
Schwarzschild black hole produces deformations which stay 
small close to the horizon but inevitably become large 
at infinity, thus 
destroying  the asymptotic flatness.

This conclusion can be supported by the following counting argument. 
Let us {\it require} the solution  to be asymptotically flat.
Then for $r\to\infty$ one should have $N=1+\delta N$, $Y=1+\delta Y$,
$U=r+\delta U$ where the variations are small. 
Inserting this into   Eqs.\eqref{eeqs} and linearizing 
with respect to $\delta N$, $\delta Y$, $\delta U$ and keeping only 
vanishing at infinity modes gives 
\bea                                      \label{infty}
N&=&1-\frac{A\sin^2\eta}{r}+B\cos^2\eta \,\frac{mr+1 }{r}\,e^{-mr},~~~~~
U=r+B\,\frac{m^2r^2+mr+1 }{m^2r^2}\,e^{-mr}, \notag \\
Y&=&1-\frac{A\sin^2\eta}{r}-B\,\sin^2\eta\,\frac{1+mr}{r}\,e^{-mr} \,,
\eea
where $A,B$ are integration constants. The other two metric amplitudes
read
\be                                         \label{infty1}
Q=1-\frac{A\sin^2\eta}{r}+\frac{2B\cos^2\eta}{r}\,e^{-mr},~~~~
a= 1-\frac{A\sin^2\eta}{r}-\frac{2B\sin^2\eta}{r}\,e^{-mr}.
\ee
This asymptotic solution is the superposition of the long-range
Newtonian mode due to the massless graviton 
and the short-range VdVZ mode describing the massive graviton  \cite{vDVZ}. 
 
Suppose that one wants to find black hole solutions with such an asymptotic 
behavior using the multiple shooting method \cite{NR}. In this method
one integrates the equations  starting from the horizon towards infinity,
and at the same time starting from infinity towards the horizon.  
The two solutions should match at some    
intermediate point,  which gives three 
matching conditions   for
$N,Y,U$. The matching conditions should be  
fulfilled by adjusting the free parameters, but   
since there are only  two parameters $A,B$ in \eqref{infty},
one cannot fulfill all three conditions. If one adjusts  also 
the parameter $u$ at the horizon, then it could be possible to 
construct global solutions, but these will  
exist at most only for discrete sets of values of $A,B,u$. 
As a result, one cannot vary $u$ continuously and so 
there could be no asymptotically flat continuous hairy deformations of
the Schwarzschild solution.

The above argument does not exclude the $U,a$ asymptotics \eqref{Ua},
since they contain altogether 5 free parameters, which is enough 
to fulfill the matching conditions.

\subsection{Hairy deformations of the asymptotically AdS black holes.}

Let us now choose in \eqref{local} 
$u=C+\delta u$ where $C=C_2$ or $C=C_3$ defined by \eqref{roots}. 
This corresponds to deformations of the asymptotically AdS black holes with 
$\Lambda(C_2)=-3.0559\,m^2$  or $\Lambda(C_3)=-1.1812\,m^2$. 
Integrating 
the equations shows that solutions with such boundary conditions exist 
 if only $|\delta u|$ is not too large. 
For $r\to\infty$ the solutions approach the corresponding background black
hole configuration \eqref{b1}, but they deviate from it
close to the horizon. The solution profiles for $C=C_3=2.6333$
are shown in Fig.2, while those for $C=C_2$  look qualitatively similar. 
Since the deformations do not change 
the asymptotic behavior  at infinity in this case, they can be viewed as
short massive hair localized in the horizon vicinity.

\begin{figure}[th]
\hbox to \linewidth{ \hss
	
	\resizebox{8.5cm}{5.2cm}{\includegraphics{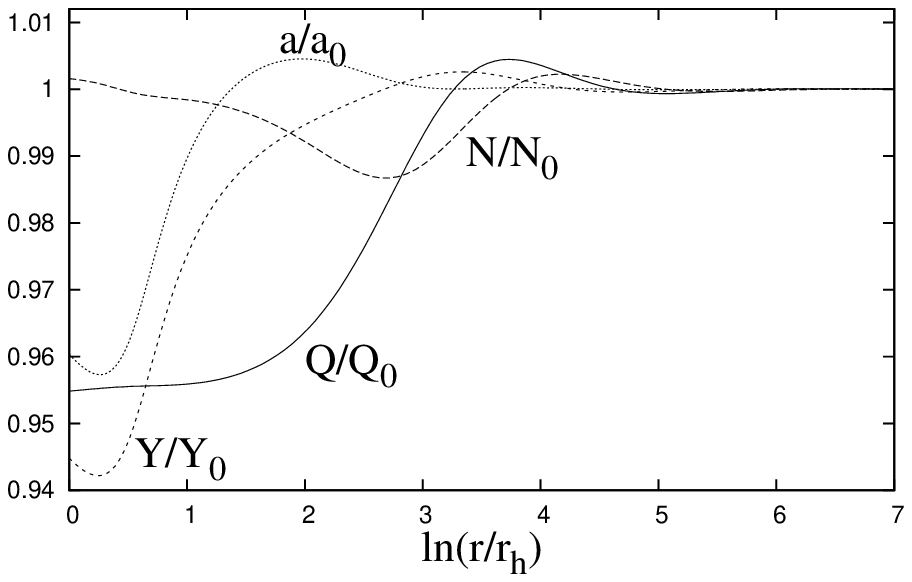}}
\hspace{1mm}
	\resizebox{8.5cm}{5.2cm}{\includegraphics{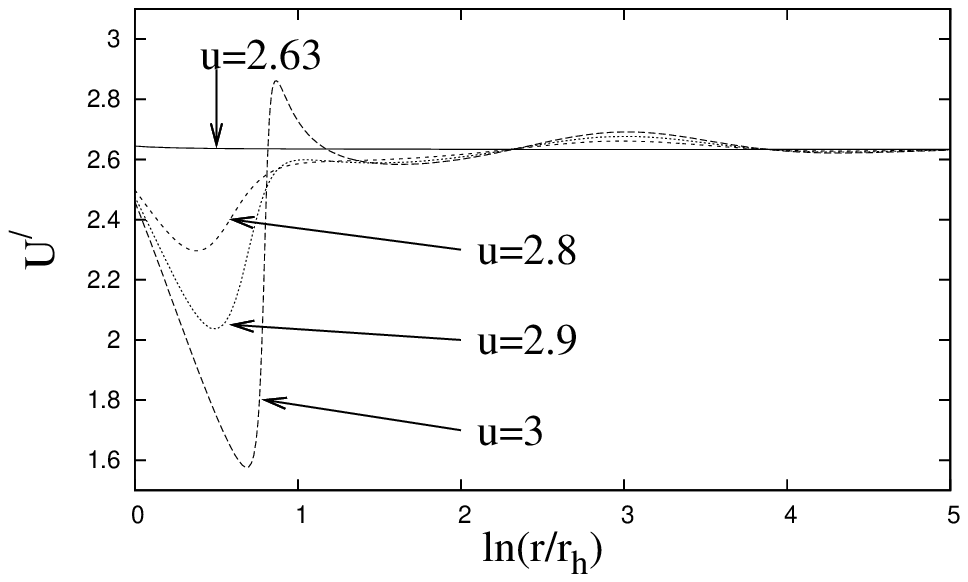}}
	
\hspace{1mm}
\hss}
\label{Fig2}
\caption{{\protect\small 
Left: profiles of the asymptotically AdS  solution
for $u=2.8$, where $N_0,Q_0,Y_0,a_0$ correspond to the 
background black hole \eqref{b1} with $C=C_3$.   
Right: $U^\prime(r)$
for several values of $u$. 
 }}%
\end{figure}

It is worth noting that, unlike for asymptotically flat solutions,  
a similar counting argument does not forbid the existence of 
continuous deformations of the asymptotically AdS black holes. 
In order to simplify the discussion, let us set $c_3=c_4=0$, in which case 
Eq.\eqref{b0} can be solved analytically, 
\be                                             \label{C000}
C=1\pm \frac{1}{\cos\eta}~~~~~\Rightarrow~~~~~\Lambda(C)=m^2(\pm\cos\eta-1).
\ee
Let $N_0,Y_0,U_0$ be the corresponding background black hole amplitudes 
and let us consider solutions that approach this background 
at infinity, $N=N_0(1+\delta N)$, $Y=Y_0(1+\delta Y)$, 
$U=U_0(1+\delta U)$ where $\delta N$, $\delta Y$, 
$\delta U$ vanish for $r\to\infty$. Inserting into \eqref{eeqs}
and linearizing gives 
\be
\delta N=\frac{A\sin^2\eta}{r^3}+O(\delta U),~~~~\delta Y=O(\delta U),~~~~
\delta U=B_{1} e^{\lambda_{1} r}+B_{2} e^{\lambda_{2} r}
\ee
with
\be
\lambda_{1}=-2+\sqrt{\frac{2\mp5\cos\eta+\cos^2\eta }{1\mp\cos\eta} },~~~~~~
\lambda_{2}=-2-\sqrt{\frac{2\mp5\cos\eta+\cos^2\eta }{1\mp\cos\eta} }. 
\ee
Since $\Re(\lambda_{1})<0$ (unless for $\eta=0,\pi$) and  $\Re(\lambda_{2})<0$,
both the $\lambda_1$ and $\lambda_2$ modes are acceptable. 
The asymptotic solution therefore 
contains tree integration constants $A,B_1,B_2$, which is enough to 
fulfill the three matching conditions within the shooting method.

\subsection{Deformations of the Schwarzschild-de Sitter black hole.}

\begin{figure}[th]
\hbox to \linewidth{ \hss

	\resizebox{8.5cm}{5.2cm}{\includegraphics{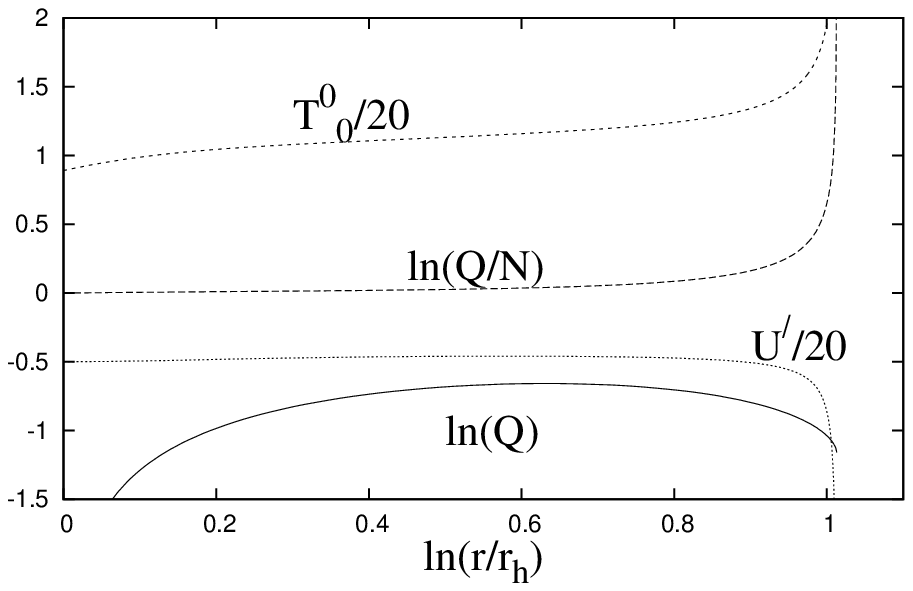}}
\hspace{1mm}


	\resizebox{8.5cm}{5.2cm}{\includegraphics{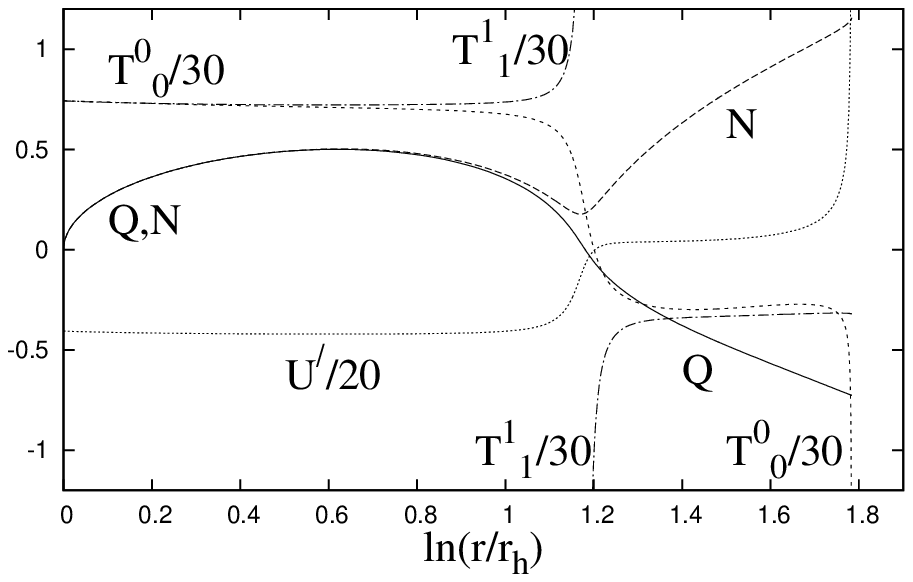}}
	
\hspace{1mm}
\hss}
\label{Fig3}
\caption{{\protect\small 
Deformations of the asymptotically dS  solution
for $u=-8$ (left) and $u=-9$ (right). 
 }}%
\end{figure}

Let us now consider the last value in \eqref{roots}, $C=C_4=-8.5566$,
which gives rise to the SdS black hole \eqref{b1}
with $\Lambda=+21.5625\,m^2$. Apart from the event horizon at $r=1$,
this solution has a cosmological horizon at $r=r_c$ where $N(r_c)=Q(r_c)=0$.
Let us set $u=C+\delta u$. 
Integrating the equations shows that
the solution becomes singular at a finite distance from the horizon. 
If $\delta u<0$ then the singularity is located at $r\approx r_c$ where
$N$ vanishes  while $Q$ does not,
while the derivatives $Q^\prime$, $(N^2)^\prime$ and  the energy 
density $T^0_0$ diverge (see Fig.3).
This implies divergence of the Riemann tensor.

Somewhat peculiar  features are shown by solutions 

with $\delta u>0$. In this case $Q$ develops a simple zero
at $r\approx r_c$ but $N$ remains finite,
so that $T^0_0$ is also finite while $T^r_r$ shows
a simple pole. The curvature diverges at this point. 
Curiously, the solution can be  continued further, 
up to a point 
where  $Q,N,T^r_r$ remain finite but $U^\prime,N^\prime$ diverge 
as does the energy density $T^0_0$ which goes to {\it minus} infinity 
(see Fig.3). 
This produces a curvature singularity, as well as an
 infinite violation of the weak energy condition. 

Summarizing, the Schwarzschild-de Sitter black hole does not admit 
non-compact, 
regular hairy generalizations, since all its deformations 
develop a curvature singularity at a finite proper distance from the horizon.  

\section{Generic black holes}
\setcounter{equation}{0}
So far we have been considering solutions which are parametrically  close to the 
background black holes, that is for $u=C_k+\delta u$ where $\delta u$ 
is not too large. Let is now consider what happens for arbitrary $u$,
when we deviate further and further away from the value $u=C_k$.

To begin with, solutions do not always exist,
since for some values of $u$ the argument of the square root in 
\eqref{a1} may become negative thus rendering the parameter
$a_1(u)$ and the local solution \eqref{local} complex-valued. 
Secondly, even if $a_1(u)$ in \eqref{a1} is real, 
it should be positive, since we assume that $N^2$ grows at $r=r_h$ 
(black hole horizon). (We do not consider the 
case where $N^2$ decreases at $r=r_h$ 
(cosmological horizon), however it can be treated by making a formal 
replacement $N\to iN$, $Q\to i$, $Y\to iY$, $a\to ia$ in the field equations.)

Having determined the allowed values of $u$,
 we integrate the equations 
and find that the solutions always reproduce one of the 
types described above. 
They approach either 
one of the $U,a$ backgrounds \eqref{Ua} so that $U^\prime(r)\to 0$ as $r\to\infty$,
or they are asymptotically $AdS$ with the cosmological constant determined
by \eqref{roots} so that $U^\prime(r)\to C_2$ or  $U^\prime(r)\to C_3$,
or they are compact and singular. 

We first study the upper branch solutions, 
with $a_1=a_1^{+}(u)$ in \eqref{local}, 
and find that they 
are non-compact only when $u$ belongs to one of the following 
regions: $I^{+}_1=[-0.53;-0.65]$, 
$I^{+}_2=[1;1.04]$, 
$I^{+}_3=[8.3;14.9]$ (see Fig.4). 
Solutions for $u\in I^{+}_1,I^{+}_2$ are the 
described above
deformations of the background AdS black hole with $U^\prime(r)=C_2$
and of the Schwarzschild solution. 
Solutions for $u\in I^{+}_3$ are new, for $r\to\infty$
they approach the background AdS black hole for $U^\prime(r)=C_3$ but cannot
be viewed as its continuous deformations, because they belong to the
different branch and so the boundary conditions at the horizon are different. 
 
For all other values of $u$ the upper branch solutions develop 
a curvature singularity at a finite proper distance from the horizon,
so that they are compact.   
In particular, the upper branch solutions seem to exist for all $u$
large and negative, but they seem to be all singular.

\begin{figure}[th]
\hbox to \linewidth{ \hss


	\resizebox{8.5cm}{5.2cm}{\includegraphics{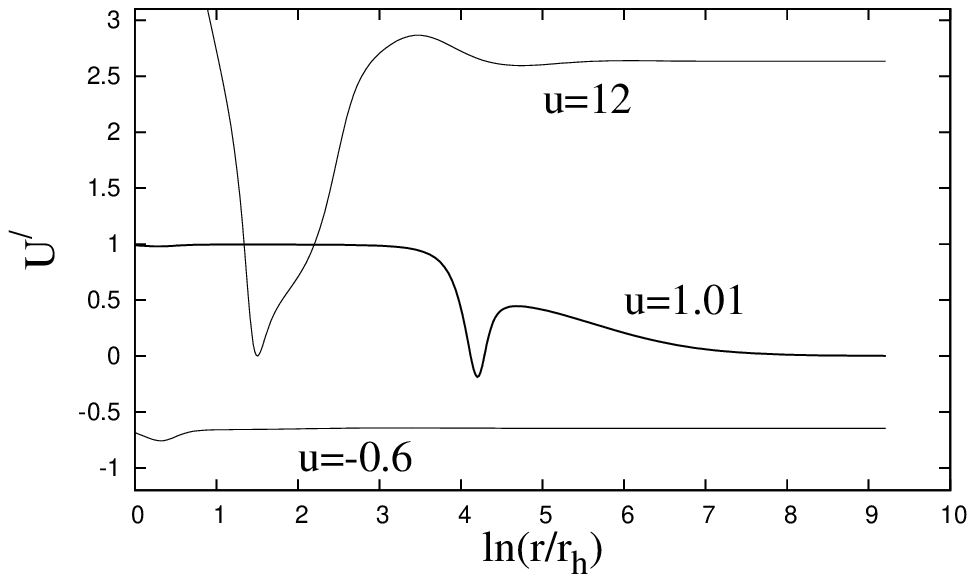}}
	
\hspace{1mm}


	\resizebox{8.5cm}{5.2cm}{\includegraphics{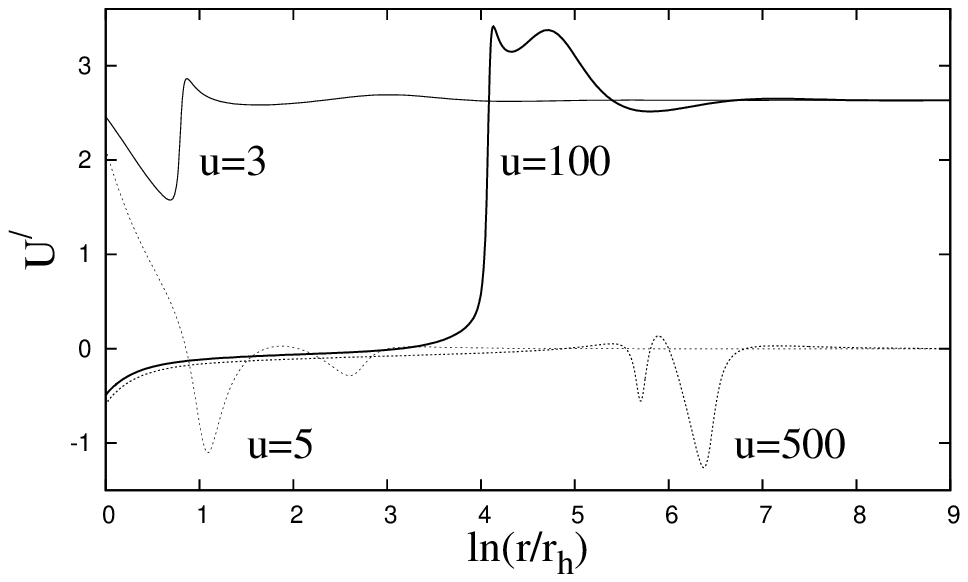}}
\hss}
\label{Fig4}
\caption{{\protect\small 
$U^\prime(r)$ for the upper (left) and lower (right) branches 
of the generic non-compact solutions. 
 }}%
\end{figure}

We then consider the lower branch solutions, 
with $a_1=a_1^{-}(u)$ in \eqref{local},  and find that they 
are non-compact if $u$ belongs to one of the four regions: 
$I^{-}_1=[2.63;3]$, 
$I^{-}_2=[3.1;29.5]$, 
$I^{-}_3=[29.9;154.2]$, 
$I^{-}_4=[154.4;1876]$. 
Solutions for $u\in I^{-}_1$ are the deformations
of the background  black hole with $U^\prime(r)=C_3$.  
For  $u\in I^{-}_2$ and $u\in I^{-}_4$ we find new asymptotically
$U,a$  solutions,
not continuously related to  the Schwarzschild metric, 
because they belong to the different branch and so have different boundary 
conditions at the horizon. 
For  $u\in I^{-}_3$ we obtain new solutions that approach the background
AdS black hole 
 with $U^\prime(r)=C_3$. It seems that for all other values of $u$ 
the solutions are compact 
and singular. 

When a solution from one branch is regular and non-compact, 
the one for the same $u$ from the second branch is usually compact and singular.  
However, this is not a general rule, since, for example, the intervals 
$I_3^{+}$ and $I_2^{-}$ have a non-zero overlap. 

The described above picture corresponds to the 
theory parameters $m=0.1$, $\eta=1$, $c_3=0.1$, $c_4=0.3$.
Varying these values changes  the size, position and  number
of intervals of $u$ within which non-compact solutions exist. 
However, the overall picture remains the same: the solutions 
either approach asymptotically the $U,a$ backgrounds 
\eqref{Ua} so that $U^\prime(r)\to 0$ as $r\to\infty$, 
or they are asymptotically $AdS$ with $U^\prime(r)\to C$
where $C$ is a root of \eqref{roots} such that $\Lambda(C)<0$, 
or they are compact and singular.
If $4c_3+c_4>6$ then the cosmological term in \eqref{Ua} 
changes sign
and the solutions that used to be asymptotically 
$U,a$ for $4c_3+c_4<6$ become compact and singular. 

The only asymptotically flat solution we find for generic
parameter values is the pure Schwarzschild black hole.
Similarly, the only asymptotically dS solution is the pure
SdS.

So far we have been assuming  the choice \eqref{al1}
for the coupling constants. The other two options  \eqref{al2} and \eqref{al3}
can be obtained by the formal complex replacements $\eta\to i\eta$ and  
$\eta\to \pi/2+i\eta$. The field equations then change but 
remain real, so that we can integrate them again. This gives new black hole solutions,
but their structure remains qualitatively the same as before -- they approach for $r\to\infty$ 
either the AdS or $U,a$ backgrounds (for the replaced value of $\eta$), or they 
are compact  and singular. We therefore find  nothing essentially new and 
so return back to the choice \eqref{al1} for the 
rest of our discussion.

\section{Special solutions}
\setcounter{equation}{0}

It is possible that new solutions
could exist for special parameter values. 
Let us see if they could be asymptotically flat. 
We know that the cosmological constant $\Lambda(C)$ in \eqref{b00} 
vanishes for $C=1$.
 However, it will also vanish 
if the expression in the parenthesis in \eqref{b00}  vanishes, that is for 
\be
C=C_{\pm}=\frac{1}{2(c_3+c_4)}\left(2c_4+5c_3-3\pm \sqrt{12c_4+9(c_3-1)^2  } \right).
\ee
Inserting this to \eqref{b3}, the polynomial $P(C)$ will vanish if either  
 $c_4=-(2/3)c_3^2$ or if $\eta=0$, in which cases we obtain additional 
background black holes 
described by the Schwarzschild metric. 
Let us now see what happens if we 
deform  these solutions by setting at the horizon $u=C_{\pm}+\delta u$.   

It turns out that nothing special  happens in the case where
$\eta$ is arbitrary and
$c_4=-(2/3)c_3^2$. There are two Schwarzschild solutions in this case, 
one for $u=C_{+}$ and another one for $u=C_{-}$. 
Choosing $u=C_{\pm}+\delta u$ produces a curvature singularity at a finite
proper distance from the horizon, similar to what is shown in 
 Fig.3. 

Let us now consider the case where $c_3,c_4$ are arbitrary but $\eta=0$. 
When $\eta$ vanishes, the source term in equation \eqref{e2} vanishes too,
so that 
the metric $f_{\mu\nu}$ becomes Ricci-flat. However, it cannot be 
flat, since our boundary conditions require that it should have a horizon,
so that it becomes the vacuum  Schwarzschild metric.
Therefore, unless we change the boundary conditions,
we do not recover for $\eta\to 0$ the RGT theory where $f_{\mu\nu}$
is non-dynamical and flat, but obtain instead the theory where 
$f_{\mu\nu}$ is non-dynamical and Schwarzschild. 
The line element \eqref{ggg1} then can be represented in the form 
\be                                \label{ggg2}
f_{\mu\nu}dx^\mu dx^\nu=a^2(U) dt^2-\frac{dU^2}{Y^2(U)}
-U^2(d\vartheta^2+\sin^2\vartheta d\varphi^2)
\ee
where 
\be                                 \label{qqq3}
Y(U)=\sqrt{1-\frac{u}{U}},~~~~~~a(U)=AY(U)
\ee
with $A$ being an integration constant. It is easy to check 
that this choice of $Y(U),a(U)$ solves equations \eqref{eq3} and \eqref{eq4}. 

Now, setting $u=C_{\pm}$ and integrating equations \eqref{eeqs}
we find that the metric $g_{\mu\nu}$ is also Schwarzschild
as it should be. 
Setting $u=C_{-}+\delta u$ gives solutions which are 
close to the Schwarzschild metric in the event horizon vicinity
but develop a curvature singularity at a finite 
distance from the horizon, so that this case is also not very interesting. 

\begin{figure}[th]
\hbox to \linewidth{ \hss

	\resizebox{8.5cm}{5.2cm}{\includegraphics{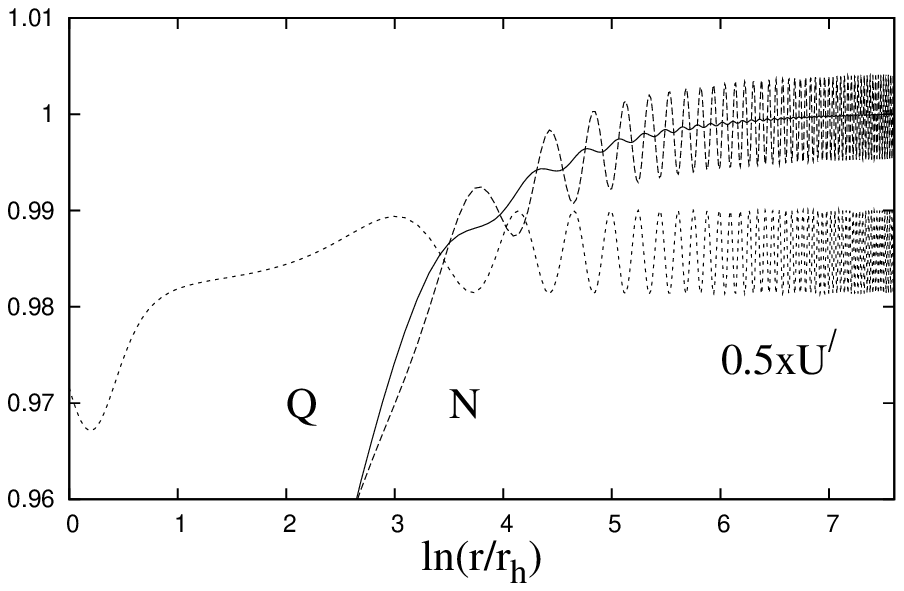}}
	
\hspace{1mm}

	\resizebox{8.5cm}{5.2cm}{\includegraphics{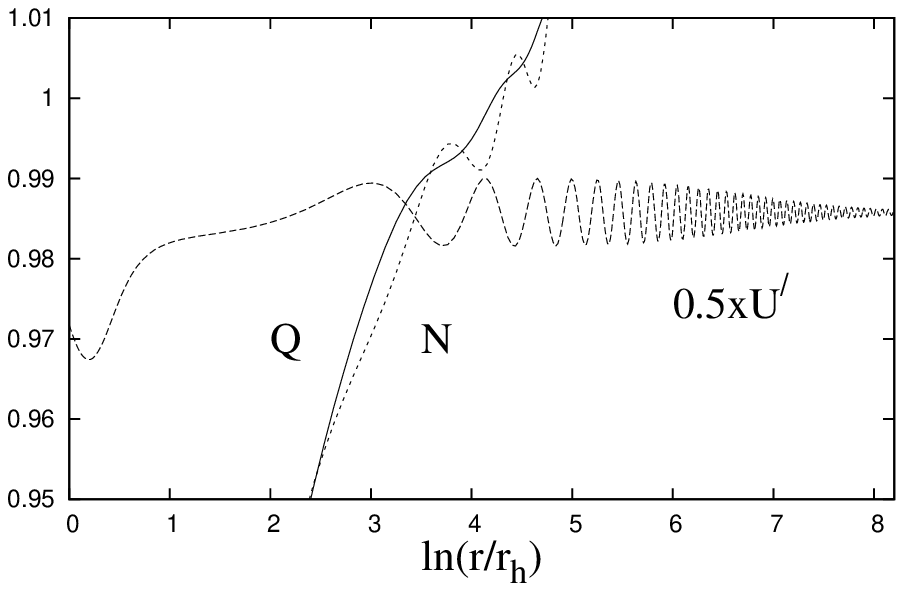}}
\hss}
\label{Fig5}
\caption{{\protect\small Solutions with 
$c_3=0.1$, $c_4=0.3$, $u=1.99$  and $\eta=0$ (left) and $\eta=0.02$
(right).   
 }}%
\end{figure}

Qualitatively 
new solutions arise for 
$u=C_{+}+\delta u$. For example,  
for $c_3=0.1$ and $c_4=0.3$ one has $C_{+}=1.97133$,
while setting $u=1.99$ gives the result shown in Fig.5.
These solutions stay always close to the background Schwarzschild solution 
but do not tend to it for $r\to\infty$ and show instead  
infinitely many  oscillations with a constant amplitude.
One has 
\be
N=\sqrt{1-\frac{1}{x}}+\delta N,~~~~
Q=\sqrt{1-\frac{1}{x}}+\delta Q,~~~~
U=C_{+}x+\delta U,~~~~
\ee
where $\delta N$, $\delta Q$, $\delta U$ are small everywhere. 
Linearizing the field equations 
gives for large $r$
\be
\delta U=\exp\{i\sqrt{2} m(x+\frac12\ln(x)) \}+\ldots,~~~~~
\delta N=-\frac{im}{\sqrt{2}}\,\delta U+\ldots,~~~~
\delta Q=\frac1x\,\delta U+\ldots,
\ee
where the real parts should be taken, 
the dots stand for the subleading terms, 
and where we have assumed for simplicity that $c_3=c_4=0$, in which 
case $C_{+}=2$. We see that the solutions behave 
as if the graviton mass was imaginary,
thus providing a
tachyonic version of the asymptotic behavior \eqref{infty},\eqref{infty1} 
obtained by linearizing 
around the $C=1$ Schwarzschild black hole. 
It is interesting that the oscillations 
persist even for finite (but small) values of $\eta$,
but the solutions become then asymptotically AdS  and 
the oscillation amplitude decreases with $r$ as shown 
in Fig.5.  
 
\section{Globally regular solutions -- stars and lumps}

For the sake of completeness, let us briefly study 
what happens when there is no horizon and the solutions are globally regular.  
They can couple to 
a compact matter source, in which case they turn out
to be asymptotically flat. If the source is absent, then the solutions describe
globally regular `lumps of self-gravitating energy'. 

\subsection{Stars}    

Let us return to equations \eqref{eq1}--\eqref{eq5}
and restore non-zero values of $\rho,P$.  
We assume the energy density to be constant inside the star
and to vanish outside, so that 
$\rho(r)=\rho_\star$ if $r<R_\star$ and  
$\rho(r)=0$ for $r>R_\star$. The pressure $P(r)$ is 
determined by equation \eqref{pressure} which should
be solved simultaneously with the other equations. 
The pressure should vanish at the surface of the star,
where the metrics $g_{\mu\nu}$ and $f_{\mu\nu}$ should be 
continuous. We require both metrics to be regular 
at the origin $r=0$, where the curvature should be 
finite.  This leads to the local 
power-series solution at small $r$ (assuming for simplicity that 
$c_3=c_4=0$)
\bea                                   \label{origin} 
N&=&1+\left(m^2\cos^2\eta\, (1-\frac{3}{2}\,u+\frac12\,u^2)
-\frac{\rho_\star}{6}\right)r^2+O(r^4),~~~~~~
U=ur+O(r^3), \notag \\
Y&=&1+m^2\sin^2\eta\,\frac{u-1}{2u}\,x^2+O(r^4),~~~~~~
P=p+O(r^2),~~~~~~Q=q+O(r^2), 
\eea
where $u,p,q$ are free parameters.  
We now wish to extend these local solutions 
numerically towards large $r$ in order to match 
the flat asymptotics \eqref{infty},\eqref{infty1}. 
Let us again count the free parameters. We need to integrate the 5 first 
order equations -- these are Eqs.\eqref{eeqs},\eqref{aaa3}
(generalized to the case of non-zero $\rho,P$) and also 
equation \eqref{pressure} for the pressure. In order to get the solutions 
within the multiple shooting method, we need 5 free
parameters in order to match the values of 5 amplitudes $N$, $Y$, $U$, $Q$, $P$
at the matching point (the amplitude $a$ is obtained afterwards from the 
algebraic constraint \eqref{aaa}). 
\begin{figure}[th]
\hbox to \linewidth{ \hss


	\resizebox{8.8cm}{5.3cm}{\includegraphics{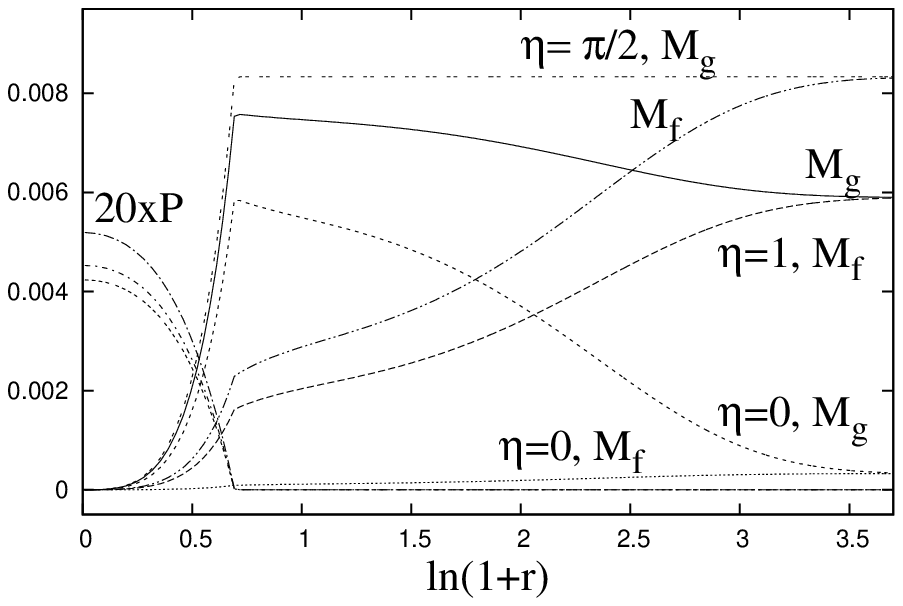}}
	
\hspace{1mm}

	\resizebox{8.8cm}{5.3cm}{\includegraphics{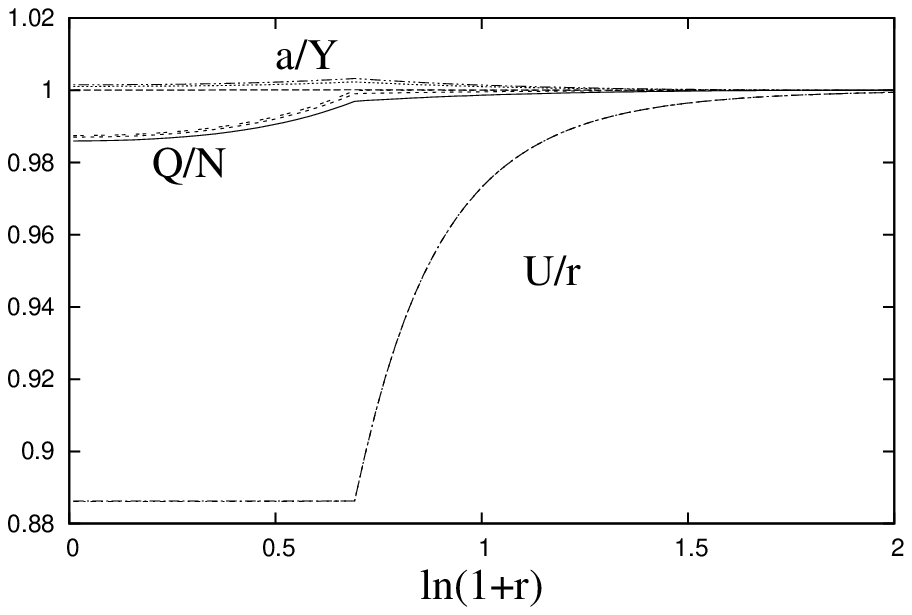}}
\hss}
\label{Fig6}
\caption{{\protect\small Star solutions for 
$c_3=c_4=0$, $m=0.2$, $\rho_\star=0.05$, $R_\star=1$ for
three different values of $\eta$. The surface of the star is located where
$P$ vanishes. 
The behavior of $U(r)$ practically does not change when $\eta$ changes,
so that the three curves for $U/r$ almost coincide and look like as one curve. 
 }}%
\end{figure}
And indeed,  
we have in our disposal exactly 5 free parameters -- these are $u,q,p$ 
in \eqref{origin} and
$A,B$ in the asymptotic solution 
\eqref{infty},\eqref{infty1}.  We can therefore do the matching,
and this gives us the global solutions. It is convenient to 
introduce mass functions $M_g$, $M_f$ 
defined by 
$N^2=1-{2M_g}/{r}$ and  $Y^2=1-{2M_f}/{U}$,
in terms of which 
the $00$-components of the Einstein equations \eqref{GGG} read
\begin{subequations}                
\begin{align}   
(M_g)^\prime &=\frac{r^2}{2}(m^2\cos^2\eta\,T^0_0+\rho),~~~~~~~\label{Mg}\\
(M_f)^\prime &=U^\prime\,\frac{U^2}{2}\, m^2\sin^2\eta\,{\cal T}^0_0. \label{Mf}
\end{align}   
\end{subequations}                
The typical solutions are shown in Fig.6. We see that inside the star, 
for $r<R_\star$, the pressure falls from its maximal value at the center till
zero at the star surface, while the mass function $M_g$ rapidly increases. 
The mass function $M_f$ also increases 
but not as fast, since it does not couple directly to $\rho$
but only to  ${\cal T}^0_0$. Outside the star $M_g$ decreases (for $\eta<\pi/2$)      
 because it 
is then sourced only by $T^0_0$ which is negative, 
while $M_f$ still increases,
since ${\cal T}^0_0$ is positive. 
For large $r$ both mass functions approach 
the same asymptotic value $M_g(\infty)=M_f(\infty)=A\sin^2\eta $
required by \eqref{infty}. 
When varying $\eta$ the parameter $A$ almost does not change, so that 
$A\sin^2\eta $
changes from the maximal value for $\eta=\pi/2$ to zero for
$\eta=0$. 

For $\eta=\pi/2$ the metric $g_{\mu\nu}$ decouples and is described by the 
Schwarzschild solution for the perfect fluid. One has in this case
\be
M_g=\rho_\star\,\frac{r^3}{6}~~~\mbox{for}~~~r<R_\star~~~~\mbox{and}~~~
M_g=M_{\rm ADM}=\rho_\star\,\frac{R^3_\star}{6}~~~\mbox{for}~~~r>R_\star
\ee 
so that the mass function $M_g$ varies only inside the star, while outside 
it is constant and equals to its asymptotic value -- the ADM mass
of the spacetime.  The  mass function $M_f$ grows 
monotonically up to the value $M_{\rm ADM}$.   

If $\eta<\pi/2$ then $M_g$ grows inside the star but not as fast as for 
$\eta=\pi/2$, since the positive contribution of $\rho$ to the right hand side
of \eqref{Mg} is partially screened by the negative $T^0_0$. Outside the star 
$T^0_0$ remains negative and continues to screen the mass of the star, so that 
$M_g$ approaches at infinity a smaller value than it has at the surface of the star. 

For $\eta=0$  the metric $f_{\mu\nu}$ decouples and becomes flat, so that $a=Y=1$, $M_f=0$,
and we recover the RGT theory (unlike for the black holes). 
The mass of the star is then
completely screened by the negative graviton energy, so that the mass function 
$M_g$ 
approaches zero at infinity.  This follows from the fact 
that the $1/r$ terms in the metric should be absent, since the 
massless graviton is `switched off' for $\eta=0$ so that the 
metric should  
approach  its asymptotic value exponentially fast.  

These solutions show the Vainstein mechanism of recovery of General Relativity 
\cite{Vainstein}. Indeed, when the graviton mass
$m$ is very small, the contribution of $m^2T^0_0$ to the 
total energy density in Eq.\eqref{Mg} becomes small as compared to $\rho$.
The mass function $M_g$ then stays approximately constant in a large region 
outside the star, in which case General Relativity is a good approximation. 

Other solutions  exhibiting the Vainstein mechanism were previously obtained 
in a massive gravity theory with the ghost \cite{Babichev} and in the RGT theory
 \cite{gruzinov}.  

\subsection{Lumps of pure gravity}
Let us now return to the local solution at the origin \eqref{origin} and set  
$\rho_\star=p=0$, in which case there is no source, and there is 
essentially only one free parameter left, $u$.
Integrating towards large $r$, we obtain a family of 
solutions labeled by $u$ and describing lumps of pure gravity. 
They have a regular center at $r=0$, while for large $r$ they 
behave similarly to the black holes and approach either the AdS backgrounds \eqref{b1},
or the $U,a$ backgrounds \eqref{Ua}, or develop a curvature singularity at a finite
proper distance from the center. 
\begin{figure}[th]
\hbox to \linewidth{ \hss


	\resizebox{8.5cm}{5.2cm}{\includegraphics{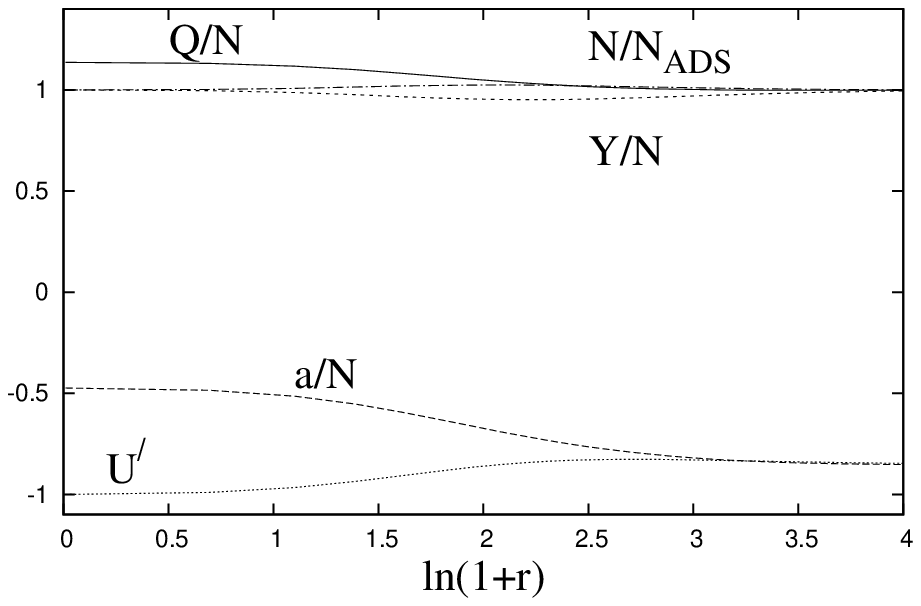}}
	
\hspace{1mm}


	\resizebox{8.5cm}{5.2cm}{\includegraphics{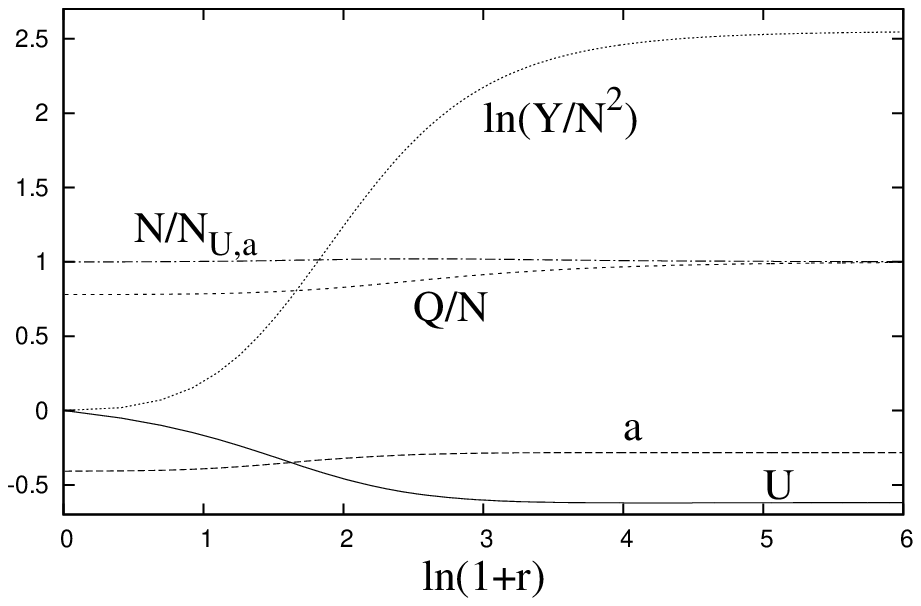}}
\hss}
\label{Fig7}
\caption{{\protect\small Lump for 
$c_3=c_4=0$, $m=0.2$, $\eta=1$ and for $u=-1$ (left)
and $u=-0.1$ (right). 
 }}%
\end{figure}
It seems that the lumps are related to the 
black holes and can be obtained form the latter in the limit when the 
event horizon shrinks. 
 Such a phenomenon is actually well known for 
hairy black holes, which can often be viewed as non-linear
superpositions of a regular gravitating matter configuration (soliton) 
with a vacuum black hole 
\cite{VG}. 

The lump shown in the left panel of Fig.7 
approaches for $r\to\infty$ the AdS background 
\eqref{b1} with $C=-0.8508$, while  
the one on the right tends to the $U,a$ background \eqref{Ua}. 
These solutions can be viewed as regular deformations of the backgrounds and
can be approximately described 
by replacing the constant $M$ in \eqref{b1},\eqref{Ua} by a function 
that approaches a constant value at large $r$ but vanishes at $r=0$,
thus insuring the regularity at the origin.

\section{Concluding remarks}
\setcounter{equation}{0}

There is something peculiar about solutions in massive gravity -- all known black holes 
are not asymptotically flat
\cite{Isham}, \cite{static}, \cite{statica}, \cite{staticb}, \cite{staticc},
but replacing the event horizon by a  regular matter source gives perfectly 
asymptotically flat systems \cite{Babichev}, \cite{gruzinov}. One can try to look for new,
possibly asymptotically flat black holes in the case where the two metrics 
are simultaneously diagonal. Since their 
event horizons should coincide in this case, this excludes  
the massive gravity theories where one of the metrics is flat \cite{DJ}. 
The same argument does not apply in the bigravity theories, 
so that new black holes could exist there. Their quest within 
the ghost-free bigravity was the subject of the above
analysis. 

We do indeed find a whole zoo of new black holes with 
massive degrees of freedom excited.  
All of them have a regular event horizon common for both metrics,
while in the outer region they show various types of behavior. 
Some of them look more conventional and support 
a short massive hair around the horizon.   
Some of them exhibit peculiar properties, as for example infinite and negative 
energy $T^0_0$ (negative energies seem to be  common for generic solutions), 
or the  tachyonic features.
However,  none of them 
 appear to be asymptotically flat. 
It seems therefore that asymptotic flatness 
is  incompatible with an event horizon 
as soon as the massive degrees of freedom are excited. 
This could perhaps be interpreted in the spirit of the no-hair theorems, 
many of which state that asymptotically flat black holes  
cannot support massive hair (see \cite{VG} for a review and references).

We have also constructed regular vacuum solutions without a horizon 
-- lumps of pure gravity. However, 
they are not asymptotically flat either  
and show in the far field the same behavior as the black holes. 
It seems that they can be obtained from the 
latter in the limit 
when the horizon shrinks to zero.

At the same time, adding a matter source we do find   
asymptotically flat solutions which exhibit the Vainstein
mechanism of 
recovery of General Relativity at finite distances. 
It seems therefore that the mechanism  needs a matter source
and does not work for pure vacuum systems like  black holes. 
This can be supported by a qualitative argument, quite in the spirit 
of Vainstein's original argumentation \cite{Vainstein}. 
If there is a matter source, then 
schematically one has 
\be
G^\rho_\lambda=m^2\, T^\rho_\lambda
+T^{{\rm (m)}\,\rho}_{~~~~\lambda} \,,
\ee
and so it is clear that if the graviton mass $m$ is very small, then 
one should be able in some way to neglect the fist term 
on the right as compared to the second one.
This reproduces General Relativity.  However, if 
$T^{{\rm (m)}\,\rho}_{~~~~\lambda}=0$, as for the studied 
above black holes or lumps, then there is no justification for omitting 
$m^2\, T^\rho_\lambda$  and General Relativity is not recovered.

\acknowledgments 

I am grateful to C\'edric Deffayet for reading the 
manuscript, remarks and suggestions, and for bringing to my attention 
Refs.\cite{staticb},\cite{DJ}.




\renewcommand{\thesection}{APPENDIX. Solutions with decoupled metrics}
\section{}
\renewcommand{\theequation}{A.\arabic{equation}}
\setcounter{equation}{0}
\setcounter{subsection}{0}

In this Appendix we analyze the 
possibility to have a non-vanishing coefficient $c$ in the 
metric $f_{\mu\nu}$ in \eqref{fff}. 
This metric is then
non-diagonal, 
which allows to find new solutions. However, these solutions are
essentially of the same type as those found long ago by 
Isham and Storey \cite{Isham},
in particular they reduce to those found in \cite{static},\cite{statica} 
when $\eta\to 0$.   

The expression $\tau^0_r$ in \eqref{T0r} must vanish, 
which is possible for $c\neq 0$
if only the expression in the parentheses vanishes. The latter condition 
requires that $U$ must be proportional to $r$,  
$U=Cr$, in which case   
\be
\tau^0_r=
\frac{c}{Q}\,\{2C-3+c_3(C^2-4C+3)+c_4(C-1)^2\},
\ee

which can be set to zero by adjusting the value of  $C$. 
Equivalently, one can consider $C,c_3$ as independent parameters,
which will be assumed below,
and this implies that 
\be
c_4=-\frac{2C-3+c_3(C^2-4C+3) }{(C-1)^2}.
\ee
This also implies that $T^0_0=T^r_r=\lambda$ with $\lambda$ given by 
Eq.\eqref{LLL} below, in which case 
the conservation condition \eqref{TTT} becomes
\be                                   \label{TTT1} 
\stackrel{(g)}{\nabla}_\mu T^\mu_r=\frac{2}{r}(T^r_r-T^\vartheta_\vartheta)=
\frac{2(c_3 C-C-c_3+2)(C^2Q-CQN b-aC+c^2N^2Q
+a b{N})}{r(C-1)Q}=0. 
\ee
Assuming for a time being that this condition is fulfilled, we shall solve the equations
and later impose it on the solutions.
If \eqref{TTT1} is fulfilled, then 
the energy-momentum tensors assume a very simple form
\be
T^\mu_\nu=\lambda\delta^\mu_\nu ,~~~~~~~~~~~~~
{\cal T}^\mu_\nu=\tilde{\lambda}\delta^\mu_\nu\,,
\ee
with 
\be                          \label{LLL}
\lambda=(C-1)(c_3C-C-c_3+3),~~~~
\tilde{\lambda}=\frac{1-C}{C^2}\,(c_3C-c_3+2), 
\ee
so that 
the field equations \eqref{e1},\eqref{e2} reduce to 
\bea
G^\mu_\nu&=&m^2\cos^2\eta\,\lambda\,\delta^\mu_\nu \,,~\label{ee1} \\
{\cal G}^\mu_\nu&=&
m^2\sin^2\eta\,
\tilde{\lambda}\,\delta^\mu_\nu    \,.       \label{ee2}
\eea
These equations describe the dynamics of the two metrics
independently driven by their cosmological terms. 
The equations for $g_{\mu\nu}$ completely decouple from those
for $f_{\mu\nu}$ so that we can solve them independently. 
With $g_{\mu\nu}$ given by
\eqref{ggg}, the solution of Eqs.\eqref{ee1} 
is the Schwarzschild-(anti)de Sitter metric,
\be
Q^2=N^2=1-\frac{2M}{r}-\frac{\lambda}{3}\,m^2\cos^2\eta\,\,r^2\,.
\ee
Let us now consider equations \eqref{ee2}
 for $f_{\mu\nu}$. They are slightly more difficult to solve,
since $f_{\mu\nu}$ is non-diagonal, while 
its components $f_{\vartheta\vartheta}=U^2$ and  
$f_{\varphi\varphi}=U^2\sin^2\vartheta$ are already fixed, since $U=Cr$.
However, the components  $f_{00}$, $f_{0r}$, $f_{rr}$ are still free, 
because they contain three up to now
unspecified functions $a,b,c$. 
We can consider $U$ as the new
radial coordinate, changing  at the same time 
the temporal coordinate, 
so that
$t\to T(t,r)$, $r\to U=Cr$. 
The metric then becomes 
\be                              \label{AdS0}
f_{\mu\nu}dx^\mu dx^\nu=f_{TT}\, dT^2+2f_{TU}dTdU+f_{UU}dU^2
-U^2(d\vartheta^2+\sin^2\vartheta d\varphi^2)\,,
\ee
where $f_{TT},f_{TU},f_{UU}$ are functions of $T,U$. 
The structure of the source term in \eqref{ee2} remains the same in the 
new coordinates, so that 
we should solve the Einstein equations with the cosmological term 
to find a metric parameterized by the radial Schwarzschild
coordinate $U$. The solution is the (anti)de Sitter metric 
\be                              \label{AdS}
f_{\mu\nu}dx^\mu dx^\nu=\Delta\, dT^2-\frac{dU^2}{\Delta}
-U^2(d\vartheta^2+\sin^2\vartheta d\varphi^2)\,,
\ee
where  $\Delta(U)=1-\frac{\tilde{\lambda}}{3}\,m^2\sin^2\eta\,U^2$. There remains 
to establish the correspondence between the $T,U$ and $t,r$ coordinates.
Let us introduce 1-forms
\bea
\theta^0=\sqrt{\Delta}dT,~~~~\theta^1=\frac{dU}{\sqrt{\Delta}},
~~~~\theta^2=Ud\vartheta,
~~~~\theta^3=U\sin\vartheta d\varphi\,,                   \label{tetrad1}
\eea 
such that 
$f_{\mu\nu}=\eta_{AB}\theta^A_\mu \theta^B_\nu$\,. 
The correspondence between the $T,U$ and $t,r$ coordinates
can be established by relating the 1-forms $\theta^A_\mu$ from \eqref{tetrad1} 
to $\omega^A_\mu$ from \eqref{tetrad}.  
The two sets of 1-forms need not coincide but may differ by a local 
Lorentz rotation. This gives two conditions 
\bea
\omega^0=\sqrt{1+\alpha^2}\theta^0+\alpha\theta^1,~~~~~~
\omega^1=\sqrt{1+\alpha^2}\theta^1+\alpha\theta^0,
\eea
where $\alpha$ is the rotation parameter. These conditions explicitly read 
\bea
a\,dt+c\,dr&=&
\sqrt{1+\alpha^2}\,(\sqrt{\Delta}\dot{T}dt+\sqrt{\Delta}T^\prime dr)
+\alpha\, \frac{Cdr}{\sqrt{\Delta}} ,   \notag \\
-cN^2\,dt+b\,dr&=&
\sqrt{1+\alpha^2}\,\frac{Cdr}{\sqrt{\Delta}}
+\alpha\,(\sqrt{\Delta}\dot{T}dt+\sqrt{\Delta}T^\prime dr ).
\eea
Comparing the coefficients in front of $dt,dr$ one finds 
\be                      \label{CT1}
T(t,r)=Ct+C\int f(r)dr
\ee
and 
\be                     \label{CT2}
a=C\sqrt{1+\alpha^2}\,\sqrt{\Delta}\,,~~~~~
b=\sqrt{1+\alpha^2}\,\frac{C}{\sqrt{\Delta}}+C\alpha\sqrt{\Delta}f\,,~~~~~
c=- C\alpha\,\frac{\sqrt{\Delta}}{N^2}
\ee
with
\be                     \label{CT3}
f=-\frac{\alpha}{\sqrt{1+\alpha^2}}\,\frac{N^2+\Delta}{N^2\Delta}\,.
\ee
In order to determine the yet unspecified function $\alpha$, we now 
use the condition that 
the expression in \eqref{TTT1} has to vanish. 
It will vanish if (since $Q=N$) 
\be                     \label{CT4}
C^2{N}-CN^2b-aC+c^2N^3 
+a b{N}=0,
\ee
which will be satisfied if 
\be
\alpha=\frac{N^2-\Delta }{2N\sqrt{\Delta } }.
\ee
Together with $U=Cr$ this finally establishes the correspondence 
between the $t,r$ and $T,U$ coordinates and   specifies all 
components of $f_{\mu\nu}$.  

The above considerations give a family of the Schwarzschild-(anti)de Sitter
backgrounds in the ghost-free bigravity for generic parameter values.   
Some additional care should be taken 
in order to make sure that all coefficients in the above expressions are real. 
For example, if the parameters are chosen such that $\lambda>0$ and
$\tilde{\lambda}<0$, then 
$\sqrt{\Delta}$ will be always real, while  $N$ will be real 
in the region between the black hole and cosmological horizons. The
solutions in regions beyond the horizons
can be obtained by the analytic continuation.  

When $\eta\to 0$ then $f_{\mu\nu}$ becomes flat 
while $g_{\mu\nu}$ does not change. 
The solutions then reduce to those obtained in the RGT theory 
\cite{static}.

Another possibility to fulfill 
Eq.\eqref{TTT1} is to restrict the value
of the coefficient $c_3$ in such a way that  
the first factor in the numerator in \eqref{TTT1} vanishes, 
\be
c_3 C-C-c_3+2=0,
\ee
which implies that 
\be
c_3=\frac{C-2}{C-1},~~~c_4=-\frac{C^2-3C+3}{(C-1)^2},~~~ 
\lambda=C-1,~~~\tilde{\lambda}=\frac{1-C}{C}. 
\ee
Since one does not need to impose the condition \eqref{CT4} in this case, 
there is no equation for the parameter $\alpha$ in 
\eqref{CT1},\eqref{CT2},\eqref{CT3} so that it remains arbitrary. 
Choosing $\alpha=0$, the $\eta\to 0$ limit of such 
solutions corresponds to the case considered in Ref.\cite{statica}.

\end{document}